\documentclass[twocolumn, tighten, numberedappendix]{aastex631}

\usepackage{amsmath,amstext}
\usepackage[T1]{fontenc}
\usepackage{booktabs}
\usepackage{apjfonts} 
\usepackage{graphicx}
\usepackage{xcolor}
\usepackage{color}
\usepackage{appendix}
\usepackage{microtype}
\usepackage{comment}
\usepackage{enumitem}
\usepackage{multirow}

\newcommand{\WMAP}{\textsl{WMAP}}

\newcommand{\wmap}{{\WMAP}}
\newcommand{\Planck}{{\textsl{Planck}}}
\newcommand{\planck}{{\textsl{Planck}}}
\newcommand{\lcdm}{\ensuremath{\Lambda}CDM}
\newcommand{\LCDM}{\ensuremath{\Lambda}CDM}
\newcommand{\kmsmpc}{\ensuremath{{\rm km\,s}^{-1}{\rm Mpc}^{-1}}}

\newcommand{\base}{\texttt{Base}}
\newcommand{\TTTEEE}{\texttt{Planck TTTEEE}}
\newcommand{\TTCut}{\texttt{Planck TT650TEEE}}
\newcommand{\Hprior}{\texttt{$\texttt{H}_\texttt{0}$}}
\newcommand{\SPT}{\texttt{SPT}}
\newcommand{\RSD}{\texttt{RSD}}
\newcommand{\DES}{\texttt{DES Y1}}

\renewcommand{\ell}{\ensuremath{l}}
\newcommand{\be}{\begin{equation}}
\newcommand{\ee}{\end{equation}}

\newcommand{\beq}{\begin{equation}}
\newcommand{\eeq}{\end{equation}}
\newcommand{\beqa}{\begin{eqnarray}}
\newcommand{\eeqa}{\end{eqnarray}}

\def\ba{\begin{eqnarray}}
\def\ea{\end{eqnarray}}

\newcommand{\barr}{\begin{array}}
\newcommand{\earr}{\end{array}}

\renewcommand{\S}{Section~}

\usepackage[flushleft]{threeparttable}
\usepackage{array}

\usepackage{savesym}
\savesymbol{tablenum}
\usepackage[separate-uncertainty=true, multi-part-units=single]{siunitx}
\restoresymbol{SIX}{tablenum}

\usepackage{gensymb}
\usepackage{float}
\usepackage{makecell}
\usepackage{hyperref}
\usepackage[hyphens]{url}

\providecommand{\sorthelp}[1]{}

\shorttitle{}

\begin{document}

\title{Cosmological Tensions and the Transitional Planck Mass Model}

\author[0000-0002-6999-2429]{Joshua A. Kable}
\affiliation{C.N. Yang Institute for Theoretical Physics, Stony Brook University \\
100 Nicolls Road, Stony Brook, NY 11790, USA \\
}

\author[0000-0002-0516-6216]{Giampaolo Benevento}
\affiliation{INFN, Sezione di Roma 2, Università di Roma Tor Vergata \\ via della Ricerca Scientifica, 1, 00133 Roma, Italy}
\affiliation{Space Science Data Center, Italian Space Agency \\ via del Politecnico snc, 00133 Roma, Italy}
\affiliation{Dipartimento di Fisica, Università di Roma Tor Vergata \\ via della Ricerca Scientifica, 1, 00133, Roma, Italy}
\affiliation{Johns Hopkins University \\
3400 North Charles Street \\
Baltimore, MD 21218, USA}


\author[0000-0002-2147-2248]{Graeme E. Addison}
\affiliation{Johns Hopkins University \\
3400 North Charles Street \\
Baltimore, MD 21218, USA}

\author[0000-0001-8839-7206]{Charles L. Bennett}
\affiliation{Johns Hopkins University \\
3400 North Charles Street \\
Baltimore, MD 21218, USA}

\begin{abstract}

In this followup analysis, we update previous constraints on the Transitional Planck Mass (TPM) modified gravity model using the latest version of EFTCAMB and provide new constraints using SPT and \planck\ anisotropy data along with \planck\ CMB lensing, BAO, SNe Ia, and an $H_0$ prior from local measurements. We find that large shifts in the Planck mass lead to large suppression of power on small scales that is disfavored by both SPT and \planck. Using only SPT TE-EE data, this suppression of power can be compensated for by an upward shift of the scalar index to $n_s = 1.003 \pm 0.016$ resulting in $H_0 = 71.94^{+0.86}_{-0.85}$ \kmsmpc\ and a $\sim7\%$ shift in the Planck mass. Including \planck\ TT $\ell \leq 650$ and \planck\ TE-EE data restricts the shift to be $<5\%$ at $2\sigma$ with $H_0 = 70.65 \pm 0.66$ \kmsmpc. Excluding the $H_0$ prior, SPT and \planck\ data constrain the shift in the Planck mass to be $<3\%$ at $2\sigma$ with a best-fit value of $0.04\%$, consistent with the \lcdm\ limit. In this case $H_0 = 69.09^{+0.69}_{-0.68}$ \kmsmpc, which is partially elevated by the dynamics of the scalar-field in the late universe. This differs from EDE models that prefer higher values of $H_0$ when high $\ell$ \planck\ TT data are excluded. We additionally constrain TPM using RSD data from BOSS DR 12 and cosmic shear, galaxy-galaxy lensing, and galaxy clustering data from DES Y1 finding both disfavor transitions close to recombination, but earlier Planck mass transitions are allowed. 

\end{abstract}

\section{Introduction} \label{intro}

The standard model of cosmology is the \lcdm\ model, which has undergone many independent tests that show good agreement with measurements from big bang nucleosynthesis (BBN) \citep{cyburt/etal:2016,Cooke/etal:2018}, primary anisotropy from the cosmic microwave background (CMB) 
\citep{bennett/etal:2013,planck/6:2018,Dutcher/etal:2021,ACTDR4}, the baryon acoustic oscillations (BAO) in galaxy distributions \citep{alam/etal:2021}, and the present accelerated expansion of the universe \citep{Brout/etal:2022}. However, \lcdm\ is not a complete model. It only provides a phenomenological description of the dark matter and the dark energy instead of a more complete physics description.

Moreover in the last decade, cosmological parameter tensions have emerged in the determination of the Hubble constant, $H_0$, and the clustering of matter, $S_8$. The $S_8$ parameter is derived from a combination of the fractional matter density, $\Omega_m$, and the amplitude of matter fluctuations on a sphere of radius 8 h$^{-1}$ Mpc, $\sigma_8$, that is given by $S_8 \equiv \sigma_8 \sqrt{\Omega_m/0.3}$.

In particular, the most precise CMB constraints today predict $H_0 = 67.37 \pm 0.54$ \kmsmpc\ assuming the \lcdm\ model \citep{planck/6:2018}, while direct measurements using the cosmological distance ladder approach using Cepheid variable stars to calibrate Type Ia Supernova find $H_0 = 73.04 \pm 1.04$ \kmsmpc\ \citep{riess/etal:2021b}. This discordance exists between several different cosmological probes, and, in general, early universe probes prefer lower values of $H_0$ while late universe probes prefer higher values of $H_0$. For a review of the $H_0$ tension see \cite{Abdalla/etal:2022}. 

Similarly, there is also a clustering tension. In particular, \planck\ predicts a value of $S_8$ assuming \lcdm\ that is up to $2-3\sigma$ higher than the values measured by cosmic shear experiments such as the Dark Energy Survey (DES), the Kilo Degree Survey (KiDS), and the Hyper Suprime-Cam (HSC) \citep{Abbot/etal:2021,heymans/etal:2021,HSC2023}. 

On the other hand, ACT DR 6 CMB lensing measurements find higher values of $S_8$, which are in agreement with the value of $S_8$ preferred by primary CMB anisotropy measurements \citep{ACTDR6Qu,ACTDR6MacCrann,ACTDR6Madhavacheril}. This restricts the space of allowed cosmological models and may hint that the $S_8$ tension may not be cleanly divided between high and low redshifts. Additionally, a reanalysis of DES Y3 and KiDS-1000 cosmic shear measurements using a consistent hybrid pipeline for both finds the combined data are in only $1.6-2.2\sigma$ tension with the \planck\ measurement \citep{DES+KiDS}.

Nevertheless, these cosmological parameter tensions may provide hints of physics beyond \lcdm, which warrants explorations. Independently, testing the effect of relaxing assumptions made by \lcdm\ in data analyses is important as a test of the \lcdm\ model. 

One of the most well-studied extensions to \lcdm\ in the last several years has been the broad class of early dark energy (EDE) models, which have been shown to alleviate the Hubble tension when fit to \planck\ data \citep{poulin/etal:2019,lin/etal:2019,Niedermann/etal:2019,karwal2021chameleon,sable/microphysics,McDonough:2021pdg, Lin:2022phm}. 

In general, EDE models involve a scalar-field that becomes dynamical around matter-radiation equality and subsequently redshifts away faster than radiation. The scalar-field has a relativistic speed of sound and increases the expansion rate of the universe prior to recombination relative to the \lcdm\ case, which results in a lowering of the size of the sound horizon at the surface of last scattering. This in turn results in an increase in the Hubble constant value preferred by CMB and galaxy BAO data \citep[see, e.g.,][for further details]{knox/millea:2020}. For a review of the impact of EDE models on the Hubble tension see \cite{DiValentino:2021izs} and \cite{Schoneberg:2021qvd}. 

There have been numerous works that have shown that EDE models can alleviate but not fully resolve the Hubble tension between \planck\ and SH0ES when a prior on $H_0$ is included. However without the $H_0$ prior, this is not the case. Moreover, EDE models tend to fit large-scale-structure (LSS) data worse than \lcdm\ because of a preference for a large increase in the cold dark matter density \citep{hill/etal:2020,ivanov/etal:2020,D'Amico/etal:2020}. 

It was shown by \cite{Vagnozzi_2020} that this results from an enhanced early integrated Sachs-Wolfe effect because of the faster than \lcdm\ expansion rate prior to recombination. To offset the enhanced decay of gravitational potential wells, more dark matter is needed, which results in increased clustering in the late universe and thus larger values of $S_8$ 

In \cite{Benevento:2022}, hereafter B22, we proposed a modified gravity model called the Transitional Planck Mass (TPM) model to address the cosmological parameter tensions. The TPM model includes a scalar-field that is coupled to gravity and becomes dynamical prior to recombination. This causes the value of the Planck mass on cosmological scales to shift, which increases the background expansion rate prior to recombination while providing a greater gravitational attraction in the early universe to hold potential wells together. 

While a time evolution of the gravitational attraction on cosmological scales is a common feature to many modified gravity models, the idea of an early time change has been explored only recently, in relation to cosmological tensions \citep[see, e.g.,][]{karwal2021chameleon,mcdonough2021early,Lin:2022phm,Sola/etal:2023}. 

In B22, we showed that a combination of \planck\ CMB anisotropy and lensing data along with BAO, Supernova, and an $H_0$ prior could provide meaningful constraints on the TPM model. In particular, the data preferred an approximately $5\%$ shift in the Planck mass some time prior to recombination. A greater gravitational attraction in the early universe was shown to be able to increase the preferred value of $H_0$ above $70$ \kmsmpc\ while allowing for $S_8$ values less than $0.8$, which alleviates both the Hubble and clustering tensions simultaneously.  

Importantly, we found that the transition in the value of Planck mass could occur over multiple decades of scale factor growth prior to recombination eliminating the coincidence problem with many EDE models, which require transitions to occur near matter-radiation equality. Additionally, we found the data were insensitive to the shape of the transition implying that the results were robust to the details of how the transition in the Planck mass would have occurred. In the late universe, the scalar-field in the TPM model behaves like a dark energy component, and thus replaces the need for a cosmological constant. However, for the case without the $H_0$ prior, we found no statistically significant preference for the TPM model over \lcdm.

In this work, we expand the analysis of the TPM model using new combinations of cosmological data. In particular, we explore the effect of using primary CMB anisotropy data from the South Pole Telescope 3G camera (SPT) instead of \planck\ anisotropy data \citep{Dutcher/etal:2021}. We also study the effect of including both the SPT and \planck\ anisotropy data. This exploration of using alternative CMB data sets is important to assess consistency of the TPM model across all measurements. 

These particular tests are additionally motivated by work that has shown that there is some preference for EDE in SPT and Atacama Cosmology Telescope (ACT) data alone and in combination with \planck\ anisotropy data where the CMB TT power spectrum is cut at $\ell = 650$ \citep{hill2021atacama,Poulin/etal:2021,Smith/etal:2022}. The multipole cut at $\ell = 650$ is done to replicate the TT power spectrum from the Wilkinson Microwave Anisotropy Probe (\wmap) \citep{bennett/etal:2013}. In \cite{Smith/etal:2022}, the full multipole range \planck\ TE and EE power spectra are also included. These tests find $H_0$ preferred values that are in better agreement with the SH0ES measurement even without requiring the SH0ES $H_0$ prior. It is therefore of interest to determine if there is a similar preference for TPM in these datasets. 

In addition to studying the effect of changing the primary CMB anisotropy data, we also explore adding more LSS data in the form of redshift space distortion (RSD) from BOSS DR12 \citep{alam/etal:2021} as well as cosmic shear, galaxy-galaxy lensing, and galaxy clustering measurements from DES Y1 \citep{des/year1:2018}. These measurements provide constraints on the clustering of matter in the universe.

In Section~\ref{Theory}, we provide a brief summary of the theory, which is fully described in B22. In Section~\ref{Data Analysis}, we outline the numerical implementation of the TPM model and the observational data sets that we include to constrain the TPM model. In Section~\ref{Result CMB} and Section~\ref{Results_LSS}, we show the results of MCMC samplings of the TPM model parameters with alternative CMB data and LSS data, respectively. Finally, in Section~\ref{Conclusions}, we provide conclusions to the results obtained and discuss potential future work.

\section{Theory}\label{Theory}

In this section, we briefly summarize the TPM model theory, which was initially introduced in B22. In this work, we make no changes to the underlying theory describing the TPM model as was described in B22 but instead expand on the data sets that we use to constrain the TPM model. Therefore, we refer the curious reader to B22 for further details of the TPM model theory. 

In the TPM model, there is a transition in the value of Planck mass in the pre-recombination universe such that the Planck mass on cosmological scales, $M_{*}^2$, is different from the present day value of the Planck mass obtained from solar system measurements, $m_0^2$.

To describe the transition in the Planck mass, we use the Effective Field Theory of Dark Energy and Modified Gravity 
(hereafter EFT) formalism \citep[see, e.g.,][for a review]{Frusciante_2019}. The TPM model is described by a function $\Omega$ that rescales the Planck mass on cosmological scales via $M_{*}^2 = m_0^2(1 + \Omega)$. In particular, the transition in the Planck mass in the TPM model is described using an error function (ERF) given by
\begin{eqnarray}
\Omega(x)=  \frac{\Omega_0}{2} \left(1- ERF \left(\frac{(x_T -x)}{\sqrt{2 \pi} \sigma}\right)\right) \label{eq:Omega} \\
\Omega'(x)=  \Omega_0 \frac{\exp{\frac{-(x-x_T)^2}{2 \sigma^2} }}{\sqrt{2 \pi} \sigma} \label{eq:Omegap},
\end{eqnarray}
where $x_T$ sets the common log of the scale factor at which the Planck mass transitions from its initial value to its present value on cosmological scales. The $\Omega_0$ parameter is the amplitude of the transition, and the $\sigma$ parameter encodes how rapid the transition in the value of the Planck mass occurs. 

Finally, the TPM model includes the parameter $c_0$, which sets the dynamics of the scalar-field in the late universe. Qualitatively, the $c_0$ parameter acts like an equation of state parameter for the scalar-field in the late universe such that more negative values of $c_0$ correspond to more negative values of the equation of state parameter. 

In some cases, we fix $c_0 = 0$. We refer to this case as the TPM f(R) model because, as discussed in B22, this case can be described using the f(R) modified gravity formalism \citep[see, e.g.,][for a review on f(R) gravity]{fR2010}.

\section{Data Analysis} \label{Data Analysis}

In this section, we outline the computational tools and techniques as well as the observational data that we use to constrain the TPM model parameters. 

\subsection{Numerical Implementation} \label{numerical}

For numerical calculations, we use EFTCAMB\footnote{\protect\url{https://github.com/EFTCAMB/EFTCAMB}} \citep{hu/etal:2014,Raveri2014}. The EFTCAMB code is a modified version of the CAMB code \citep{Lewis_2000} that includes the EFT description of modified gravity and dark energy. In B22, the TPM model was implemented in EFTCAMB.

To sample the posterior distribution functions for the model parameters, we run Markov Chain Monte Carlo (MCMC) using EFTCosmoMC. The EFTCosmoMC code is built on the existing CosmoMC code \citep{Lewis_2002}, but uses EFTCAMB instead of CAMB. To assess the convergence of the MCMC chain, we use a Gelman-Rubin convergence statistic of $R-1 = 0.01$ \citep{gelman/rubin:1992}. For each MCMC, we use flat priors on the TPM model parameters given by $-7 \leq x_T \equiv \textrm{log}(a_T) \leq -3$, $0.1 \leq \sigma \leq 3$, $-0.15 \leq \Omega_0 \leq 0.01$, and $-0.04 \leq c_0 \leq 0.1$. 

We determine the best-fit parameter values for all models using the BOBYQA algorithm \citep{Powell} included in EFTCosmoMC and CosmoMC. For the starting point of the BOBYQA algorithm, we use the mean values from the MCMCs for each of the parameters. 

We note that since the publication of the first TPM paper, B22, there have been updates to the EFTCAMB code that result in shifts in parameters at the $1 \sigma$ level for both the TPM and \lcdm\ models. We compared the ability of the combination of EFTCAMB and EFTCosmoMC to recover the posterior distributions of parameters to the ability of the latest versions of CAMB and CosmoMC when assuming \lcdm. We replaced the \Planck\ 2018 TTTEEE $30 < \ell \leq 2508$ data with a best-fit theory model from the \lcdm\ fit to the \Planck\ 2018 likelihood. For these power spectra, we do not include any sources of noise or cosmic variance. We ran MCMCs using both EFTCosmoMC and CosmoMC and find that in each case we recover the input \lcdm\ parameters to within $0.1 \sigma$. This implies that there is good agreement between the new versions of EFTCAMB/EFTCosmoMC and the new versions of CAMB/CosmoMC. 

In all cases, we only solve the evolution of linear cosmological perturbations, as we do not currently have a prescription for the non-linear growth of matter perturbations in the TPM model. 

\subsection{Observational Data}\label{Data}

In this subsection, we provide a summary of the data sets that we use to constrain the TPM model in this work. In B22, the TPM model was constrained by a Baseline likelihood that included \planck\ primary anisotropy and CMB lensing data as well as BAO, Supernova, and a local prior on $H_0$. 

In this work, we expand on this analysis and in particular explore the effects of replacing \planck\ primary anisotropy data with SPT 3G primary ansiotropy data as well as combining both data sets. In addition, we also constrain the TPM model using RSD and weak lensing measurements.

In the remainder of this subsection, we outline what data sets are used in this analysis. In all cases we include

\begin{itemize}
    \item \textbf{\texttt{BAO}:} BAO data from BOSS DR 12 \cite{alam/etal:2017} as well as the SDSS Main Galaxy Sample \citep{ross/etal:2015} and the 6dFGS survey \citep{beutler/etal:2011}.
     \item \textbf{\texttt{Planck Lensing}:} \planck\ 2018 lensing likelihood \citep{PlanckCollaboration:2020b}.
    \item \textbf{\texttt{Supernova}:} Pantheon 2018 compilation of Type Ia supernova that does not include an absolute calibration \citep{scolnic/etal:2018}.
\end{itemize}

We refer to this combination of data as the \texttt{Base} likelihood. These data sets were all included in the B22 TPM analysis. In addition to the \texttt{Base} data combination, we also include various combinations of SPT and \planck\ CMB primary anisotropy data. In particular, we will refer to these combinations as

\begin{itemize}
    \item \textbf{\texttt{SPT}:} Measurements of the TE and EE power spectra from SPT 3G over the multipole range $300 \leq \ell < 3000$ \citep{Dutcher/etal:2021}. For cases where we do not include \planck\ Low $\ell$ EE data, we include a prior on $\tau = 0.054 \pm 0.0073$.
    \item \textbf{\texttt{Planck TTTEEE}: }  Plik Lite 2018 TTTEEE along with \Planck\ TT $\ell \leq 30$, and \planck\ Low $\ell$ EE data \citep{PlanckCollaboration:2020b}. 
    \item \textbf{\texttt{Planck TT650TEEE}:}  Subsets of the Plik Lite 2018 likelihood data sets including multipole cuts in TT at $\ell \leq 650$. This approximates the constraints provided by \wmap\ TT data \citep{bennett/etal:2013}. Additionally, we include \Planck\ TT $\ell \leq 30$, and \planck\ Low $\ell$ EE data \citep{PlanckCollaboration:2020b}.
\end{itemize}

Additionally we explore the ability of RSD and weak lensing measurements to constrain the TPM model. We will refer to these data as

\begin{itemize}
    \item \textbf{\texttt{RSD}:} Redshift space distortion measurements from BOSS DR 12 \citep{alam/etal:2017}. These measurements are correlated with the other BOSS DR 12 BAO data meaning a combined BAO + RSD likelihood replaces the \texttt{BAO} likelihood above. 
    \item \textbf{\texttt{DES Y1}:} Measurements from the Dark Energy Survey Y1 cosmic shear, galaxy-galaxy lensing, and galaxy clustering \citep{des/year1:2018}. These data sets are collectively referred to as a 3x2 point analysis. We use the aggressive linear scale cuts in all cases \citep[see][for details]{des/year1_extended}. 
\end{itemize}

Finally, we explore the effect that including a prior on $H_0$ has on the allowed parameter space for the TPM model. This prior is given by

\begin{itemize}
    \item \textbf{$\texttt{H}_\texttt{0}$:} A prior given by $H_0 = 72.61 \pm 0.89$ \kmsmpc\ corresponding to a combination of local measurements \citep{riess/etal:2021b,Pesce2020,Blakeslee2021}.
\end{itemize}

For some data combinations explored in this analysis, we combine \planck\ and SPT anisotropy data. For the cases where we combine \planck\ and SPT data, we make a multipole cut on \planck\ TT at $\ell \leq 650$\footnote{For the \planck\ multipole cuts, we artificially increase the uncertainties in the \planck\ data covariance matrix for TT $\ell > 650$ for the TT-TT, TT-TE, and TT-EE covariance blocks. We check that this method results in a $\Delta \chi^2$ with respect to the true $\chi^2$ of less than $0.001$ for the TPM model meaning we are recovering the correct $\chi^2$ to within this error.}. \planck\ TT $\ell \leq 650$ provides a good approximation for combining SPT data with \wmap\ data as has been explored in \cite{hill2021atacama} and \cite{Smith/etal:2022}. Including a \wmap-like dataset provides a large angular scale anchor that complements SPT data. 

Additionally, we include the full multipole range \planck\ TE and EE data to see how adding additional polarization data can tighten constraints on the TPM model. To a good approximation, the covariance between \planck\ TE or EE and SPT TE or EE is negligible because of noisier \planck\ polarization measurements, and the fact that SPT sky coverage comprises a small fraction of the total area covered by the \planck\ measurements \citep{Dutcher/etal:2021,Balkenhol/etal:2022}. 

For weak lensing measurements, we include cosmic shear, galaxy-galaxy lensing, and galaxy clustering measurements from DES Y1. In all cases, we use the DES Y1 built in aggressive linear scale cuts because we do not currently have accurate modelling of the non-linear collapse of dark matter on small scales \citep{des/year1_extended}. We note, however, that these scale cuts were made by the DES collaboration assuming the \lcdm\ model and the scale where non-linearities become relevant may in general be different for the TPM model.

\section{Using SPT 3G Anisotropy Data to Constrain the TPM Model} \label{Result CMB}

In this section, we study the effect of fitting the TPM model to SPT primary anisotropy data instead of \planck. In particular, we use data from the SPT-3G camera, which operates from the South Pole and covers a survey field of approximately 1500 deg$^2$ patch of the sky. In Section~\ref{SPT_only}, we show the differences between using SPT versus \planck\ on the constraining power of the \lcdm\ and TPM parameters. In Section~\ref{SPT_Planck}, we show parameter constraints from a combination of SPT and \planck\ anisotropy data. In Section~\ref{SPT_discussion}, we provide discussions on the differences between the preferred TPM parameter values when fit to SPT or \planck\ data. For supplementary material related to the TPM model fits to each data sets, in Appendix~\ref{SPT_H0}, we show the effect of including/excluding a prior on $H_0$, and in Appendix~\ref{SPT_fR}, we show the effect of fixing $c_0 = 0$, which corresponds to the TPM f(R) case. 

\subsection{Fits to Either SPT or \Planck\ Anisotropy Data} \label{SPT_only}

\begin{figure*}[!tbp]
\centering

\hspace{-1cm}
\includegraphics[clip,width=\linewidth]{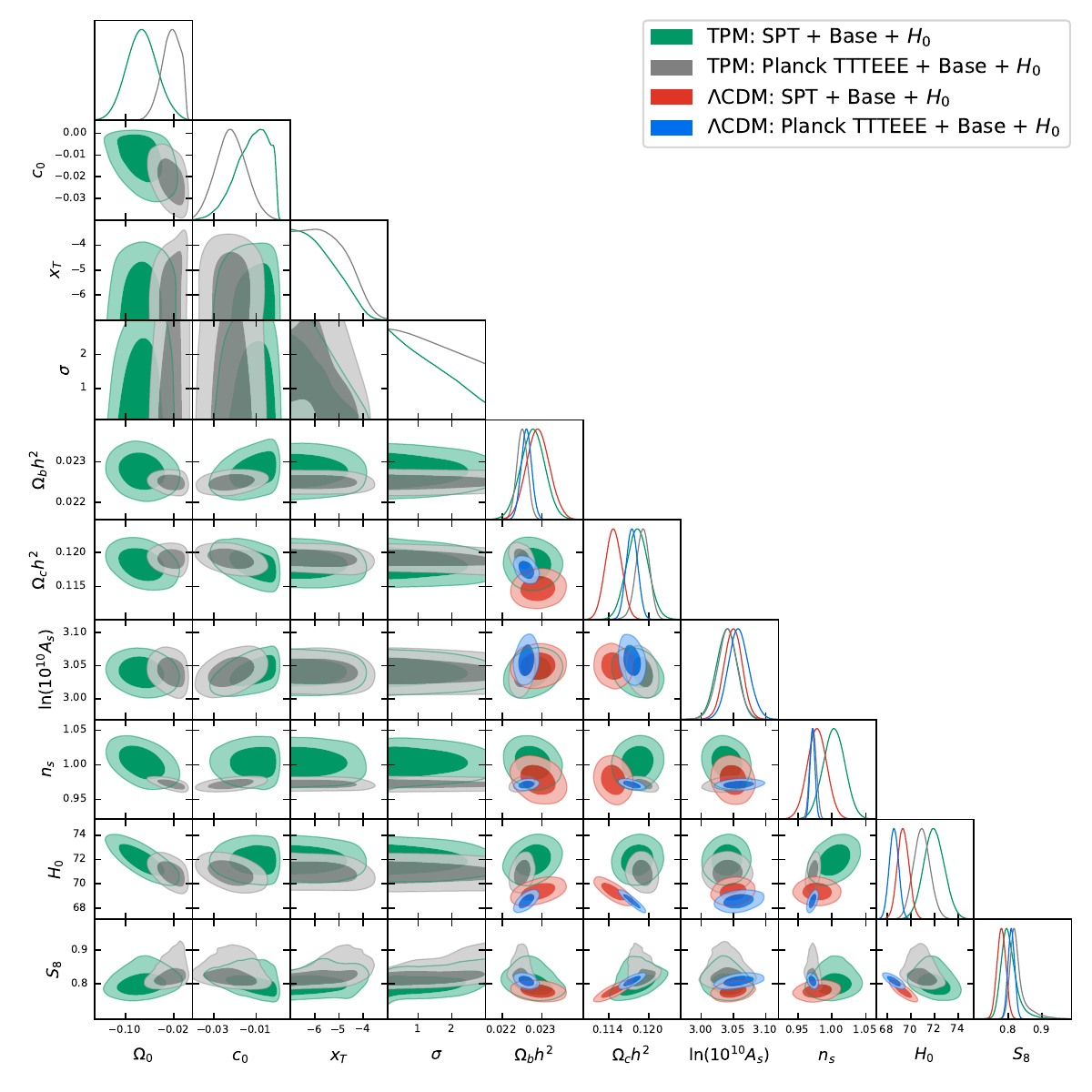}

\vspace{-3mm}

\caption{ The \SPT\ + \base\ + \Hprior\ (green) combination are less constraining of the TPM model than the \TTTEEE\ + \base\ + \Hprior\ (gray) combination, where the \base\ dataset includes \planck\ CMB lensing, BAO, and Supernova data. In particular, broader posterior distributions for $\Omega_bh^2$, $\Omega_ch^2$, and $n_s$ allow for larger amplitude (more negative) transitions in the Planck mass with a best-fit value of $\Omega_0 = -0.073$ or about a 7.3$\%$ shift in the Planck mass. This shift in the Planck mass is compensated for by a large increase in $n_s$ ($n_s = 1.003 \pm 0.016$) and allows for higher values of $H_0$ ($H_0 = 71.94^{+0.86}_{-0.85}$ \kmsmpc) and lower values of $S_8$ ($S_8 = 0.802^{+0.022}_{-0.023}$) than the TPM fit to \TTTEEE\ + \base\ + \Hprior. Because the \planck\ mass is allowed larger amplitude transitions, there is less need for shifts in the $c_0$ parameter, which acts as an equation of state parameter for the dark energy in the late universe, to raise $H_0$ to better match the $H_0$ prior. This results in a preferred value of $c_0$ closer to $0$. In both SPT and \planck\ cases, the timing of the transition, $x_T$, and the rapidity of the transition, $\sigma$, are unconstrained by the data. Notably, the \SPT\ + \base\ + \Hprior\ constraints of $S_8$ from \lcdm\ (red) are lower than the corresponding case for TPM suggesting that the TPM model is not doing better on clustering than \lcdm\ for this data combination. 
}

\label{fig:Comparison_SPT_no_plik}
\end{figure*}

In this subsection, we compare how the \lcdm\ and TPM models fit to data combinations that include either SPT or \planck\ anisotropy data. In particular, we run MCMCs for both models in fits to \texttt{Planck TTTEEE} + \texttt{Base} + $\texttt{H}_\texttt{0}$ and \texttt{SPT} + \texttt{Base} + $\texttt{H}_\texttt{0}$ (see Section~\ref{Data} for descriptions of each of these data sets). We show the resulting parameter constraints from the MCMCs in Table~\ref{tab:parameters_SPT_no_plik} and Figure~\ref{fig:Comparison_SPT_no_plik}. 

\begin{table*}[!tbp]
\setlength{\tabcolsep}{5pt}
\centering
\hspace{-1cm}
\begin{tabular}{@{}cccccc@{}}


\toprule

 & $\Lambda$CDM: \TTTEEE\                        & $TPM$: \TTTEEE\                      & $\Lambda$CDM: \SPT\   & $TPM$: \SPT\  \\ 
& + \base\ + \Hprior\ & + \base\ + \Hprior\ & + \base\ + \Hprior\ & + \base + \Hprior\ \\
\toprule                                                                                                               
$100\theta_{\rm MC}$   & 1.0413 ($1.04127^{+0.00029}_{-0.00030}$)   & 1.04139 (1.04135 $\pm$ 0.00036)  & 1.03992 ($1.03984^{+0.00064}_{-0.00065}$) & 1.04039 ($1.04042^{+0.00071}_{-0.00072}$ )  \\
$\Omega_bh^2$          & 0.02261 (0.02261 $\pm$ 0.00013)   & 0.022505 (0.022498 $\pm$ 0.00013)  & 0.02288 (0.02290 $\pm$ 0.00029) & 0.02271 (0.02277 $\pm$ 0.00030)  \\
$\Omega_ch^2$          & 0.11748 (0.11748 $\pm$ 0.00086)   & 0.11934 (0.11906 $\pm$ 0.00099)  & 0.1148 (0.1147 $\pm$ 0.0012) & 0.1182 (0.1183 $\pm$ 0.0016)     \\
$\tau$                 & 0.0615 ($0.0633^{+0.0080}_{-0.0079}$)   & 0.0532 (0.0528 $\pm$ 0.0074 )  & 0.0556 (0.0563 $\pm$ 0.0070) & 0.0610 (0.0543 $\pm$ 0.071)        \\
$\ln(10^{10}A_s)$      & 3.054 ($3.058^{+0.016}_{-0.015}$)   & 3.043 (3.040 $\pm$ 0.015)  & 3.047 (3.049 $\pm$ 0.014 ) & 3.049 (3.039 $\pm$ 0.016)        \\
$n_s$                  & 0.9712 (0.9713 $\pm$ 0.0036)   & 0.9715 (0.9721 $\pm$ 0.0048)  & 0.978 (0.978 $\pm$ 0.014) & 1.008 (1.003 $\pm$ 0.016)     \\
\midrule

$\Omega_0$                  & -   & -0.025 (> -0.058 at 95$\%$)  & - & -0.073 (-0.072 $\pm$ 0.025)      \\
$x_T$         & -   & -5.33 (-5.58 $\pm$ 0.99)  & - & -5.06 ($-5.81^{+0.91}_{-0.86}$)         \\
$\sigma$         & -   & 0.82 ($1.42^{+0.98}_{-0.93}$)  & - & 1.35 ($1.23^{+0.90}_{-0.84}$)        \\
$c_0$              & -   & -0.02293 (-0.02174 $\pm$ 0.0071) & - & -0.0111 (> -0.0287 at 95$\%$)           \\
\midrule
$H_0$                  & 68.56 (68.57 $\pm$ 0.39)   & 70.94 ($70.90^{+0.69}_{-0.70}$ )   & 69.32 ($69.34^{+0.48}_{-0.47}$) & 71.87 ($71.94^{+0.86}_{-0.85}$)       \\
$\sigma_8$                  & 0.8076 (0.8091 $\pm$ 0.0064)   & 0.839 ($0.853^{+0.020}_{-0.021}$) & 0.7948 ($0.7952^{+0.0076}_{-0.0077}$) & 0.830 ($0.840^{+0.020}_{-0.021}$)         \\
$S_8$                  & 0.8068 ($0.8083^{+0.0098}_{-0.0099}$)   & 0.815 ($0.828^{+0.020}_{-0.021}$)  & 0.778 (0.779 $\pm$ 0.013) & 0.794 ($0.802^{+0.022}_{-0.023}$)        \\
$\chi^2_{\rm Planck TTTEEE}$ & 589.78   & 585.69 & - & -                 \\
$\chi^2_{\rm Planck low TT}$ & 22.41   & 21.41  & - & -                \\
$\chi^2_{\rm Planck lowE}$ & 397.70   & 395.83  & - & -                  \\
$\chi^2_{\rm SPT}$ & -   & -  & 1120.67 & 1120.67                    \\
$\chi^2_{\rm CMB \ lensing}$ & 9.575   & 8.82  & 8.95 & 9.17                \\
$\chi^2_{\rm BAO}$ & 5.55   & 7.74  & 8.27 & 6.32                     \\
$\chi^2_{\rm Pantheon}$ & 1034.73   & 1037.08  & 1035.00 & 1035.23                     \\
$\chi^2_{\rm H_0}$ & 20.68   & 3.52  & 13.66 & 0.69                   \\
$\chi^2_{\rm prior}$ & 0.22   & 0.09  & 0.66 & 0.41                   \\
$\chi^2_{\rm tot}$ & 2081.39   & 2060.19  & 2187.20 & 2172.48                     \\
\end{tabular}
\caption{ \label{tab:parameters_SPT_no_plik} Maximum Likelihood Parameter Values and, in Parenthesis, Mean plus 68$\%$ Confidence Level Bounds. Each Column Delineates the Model and Likelihood Combination (See Section~\ref{Data} for Details of the Likelihoods).
}
\end{table*}

\begin{table}[!tbp]
\setlength{\tabcolsep}{5pt}
\centering
\begin{tabular}{cc}

\toprule
Parameter & Shift ($\sigma$) \\
\toprule                                                                                                  
$100\theta_{\rm MC}$   & -0.5 $\sigma$  \\
$\Omega_bh^2$          & -0.9 $\sigma$  \\
$\Omega_ch^2$          & 0 $\sigma$ \\
$\tau$                 &  -0.2 $\sigma$ \\
$\ln(10^{10}A_s)$      & -0.4 $\sigma$   \\
$n_s$                  & -0.6 $\sigma$ \\
\midrule
$\Omega_0$                  & 1.3 $\sigma$   \\
$x_T$         &  -0.3 $\sigma$ \\
$\sigma$         &    0.7 $\sigma$   \\
$c_0$              &  -0.6 $\sigma$ \\
\midrule
$H_0$                  &   -0.6 $\sigma$    \\
$\sigma_8$                  & 0 $\sigma$  \\
$S_8$                  & 0 $\sigma$  \\
\end{tabular}
\caption{ Shifts in the Mean Value of Each Model Parameter from the TPM Fit to \TTTEEE\ + \base\ + \Hprior\ in B22 to the Mean Value of Each Model Parameter Fit to the Same Likelihood in This Work. All Shifts Calculated Using Uncertainty in Table~\ref{tab:parameters_SPT_no_plik} \label{tab:shift_in_parameters} ).
}
\end{table}

The \TTTEEE\ + \base\ + \Hprior\ was the Baseline data combination explored in B22. As noted in Section~\ref{numerical}, since the publication of B22, there have been some improvements in the EFTCAMB code that have resulted in shifts in parameter values. We find that these changes bring the EFTCAMB code into better agreement with the more recent CAMB releases. We report all of these shifts in model parameters in Table~\ref{tab:shift_in_parameters}. The largest shift is a 1.3$\sigma$ higher (less negative) value for $\Omega_0$ so that the data prefer approximately a $2.5\%$ shift in the Planck mass. Because the $H_0$ prior penalizes lower values of $H_0$ and smaller shifts in the Planck mass result in lower preferred values of $H_0$, the $c_0$ parameter decreases (becomes more negative) relative to the value reported in B22 by $0.6\sigma$ to increase the preferred value of $H_0$.

The combination of these changes still leads to a shift downward in $H_0$ relative to B22 by $0.6\sigma$ to $H_0 = 70.90^{+0.69}_{-0.70}$ \kmsmpc. Thus the TPM model still reduces the Hubble tension between \planck\ CMB data and local measurements. 

The TPM model fits the \SPT\ + \base\ + \Hprior\ data better than \lcdm. Quantitatively, the overall $\Delta \chi^2 = -14.72$, though most of this improvement comes from the increase in the preferred value of $H_0$ ($\Delta \chi^2 = -12.97$). We explore the effect of not including the $H_0$ prior in Appendix~\ref{SPT_H0}. To summarize, we find that the TPM model fit to \SPT\ + \base\ data has parameter posteriors with non-Gaussian tails that allow for $H_0 > 70$, $S_8 < 0.80$, and $\Omega_0 < -0.05$, though these parameter values are not the peaks of the posteriors. 

The improvement found by the TPM model fit to \SPT\ + \base\ + \Hprior\ compared to the \lcdm\ fit to the same data is smaller than the corresponding improvement found by the TPM model fit to \TTTEEE\ + \base\ + \Hprior\ compared to the \lcdm\ fit, $\Delta \chi^2 = -21.20$. 

Relative to the TPM fit to \TTTEEE\ + \base\ + \Hprior, the TPM fit to \SPT\ + \base\ + \Hprior\ have broader posteriors for $n_s$, $\Omega_bh^2$, and $\Omega_ch^2$, which are shown in Figure~\ref{fig:Comparison_SPT_no_plik}. Notably, these uncertainties are only marginally larger than the uncertainties on the corresponding parameters in the \lcdm\ fit to \SPT\ + \base\ + \Hprior. 

These larger uncertainties on standard \lcdm\ parameters allow for more negative transitions in the Planck mass so that \SPT\ + \base\ + \Hprior\ prefers $\Omega_0 = -0.072 \pm 0.025$, which corresponds to a nonzero shift in the Planck mass at $2.9\sigma$. As a result, the constraint on $H_0$ shifts to $H_0 = 71.94^{+0.86}_{-0.85}$ \kmsmpc, which corresponds to a $\chi^2 = 0.69$ with the $H_0$ prior constraint from local measurements resolving the Hubble tension between these specific data. Note that this preference for a shift in the Planck mass is driven by the prior on $H_0$. 

These parameter shifts in $\Omega_0$ and $H_0$ are compensated for by an increase in the scalar index to $n_s = 1.003 \pm 0.016$, which is notably larger than the value of $n_s$ preferred by the \lcdm\ fit to \SPT\ + \base\ + \Hprior. The TPM fit to \SPT\ + \base\ + \Hprior\ also has a $0.9\sigma$ shift upward in $\Omega_bh^2$ and $1.3\sigma$ shift downward in $100\theta_{MC}$ relative to the TPM fit to \TTTEEE\ + \base\ + \Hprior. While the best-fit value of $S_8$ is lower for the TPM fit to \SPT\ + \base\ + \Hprior\ than the TPM fit to \TTTEEE\ + \base\ + \Hprior, the best-fit $S_8$ value for the \lcdm\ fit to \SPT\ + \base\ + \Hprior\ is lower than both. This lower preferred values of $S_8$ for \lcdm\ fits to SPT is consistent with the results obtained in \cite{Dutcher/etal:2021}. 

For the remainder of the TPM model parameters, both the scale factor of the transition in the Planck mass, quantified by $x_T$, and the rapidity of the transition, $\sigma$, are poorly constrained by both \planck\ and SPT anisotropy data. This implies that: 1) the transition is free to occur over several decades of scale factor, which removes the so-called "Why now?" problem with many EDE models, and 2) these results are robust to changes in the phenomenology of how the transition occurs. 

Finally, the preferred value of the $c_0$ parameter is lower for the TPM fit to \SPT\ + \base\ + \Hprior\ than the TPM fit to \TTTEEE\ + \base\ + \Hprior. The $c_0$ parameter acts like an equation of state parameter for the scalar field in the late universe. Lowering the value of $c_0$ leads to an increase in $H_0$, so because the $\Omega_0$ parameter already increases $H_0$ to be in agreement with the $H_0$ prior, there is less need for the $c_0$ parameter to be nonzero.   

To summarize, we find that \textit{SPT anisotropy data have a broader allowed parameter space than the \planck\ anisotropy data, which allows for larger shifts in $H_0$ that resolve the Hubble tension between SPT and SH0ES. These shifts are compensated for by a shift upward in the scalar index to $n_s > 1$. While these shifts in the preferred values of $H_0$ and $\Omega_0$ are driven by the $H_0$ prior, this part of parameter space is allowed within $95\%$ confidence intervals even when the $H_0$ prior is excluded.}

\subsection{Fits to Combined SPT and Planck Anisotropy Data} \label{SPT_Planck}

\begin{figure*}[!tbp]
\centering

\hspace{-1cm}
\includegraphics[clip,width=\linewidth]{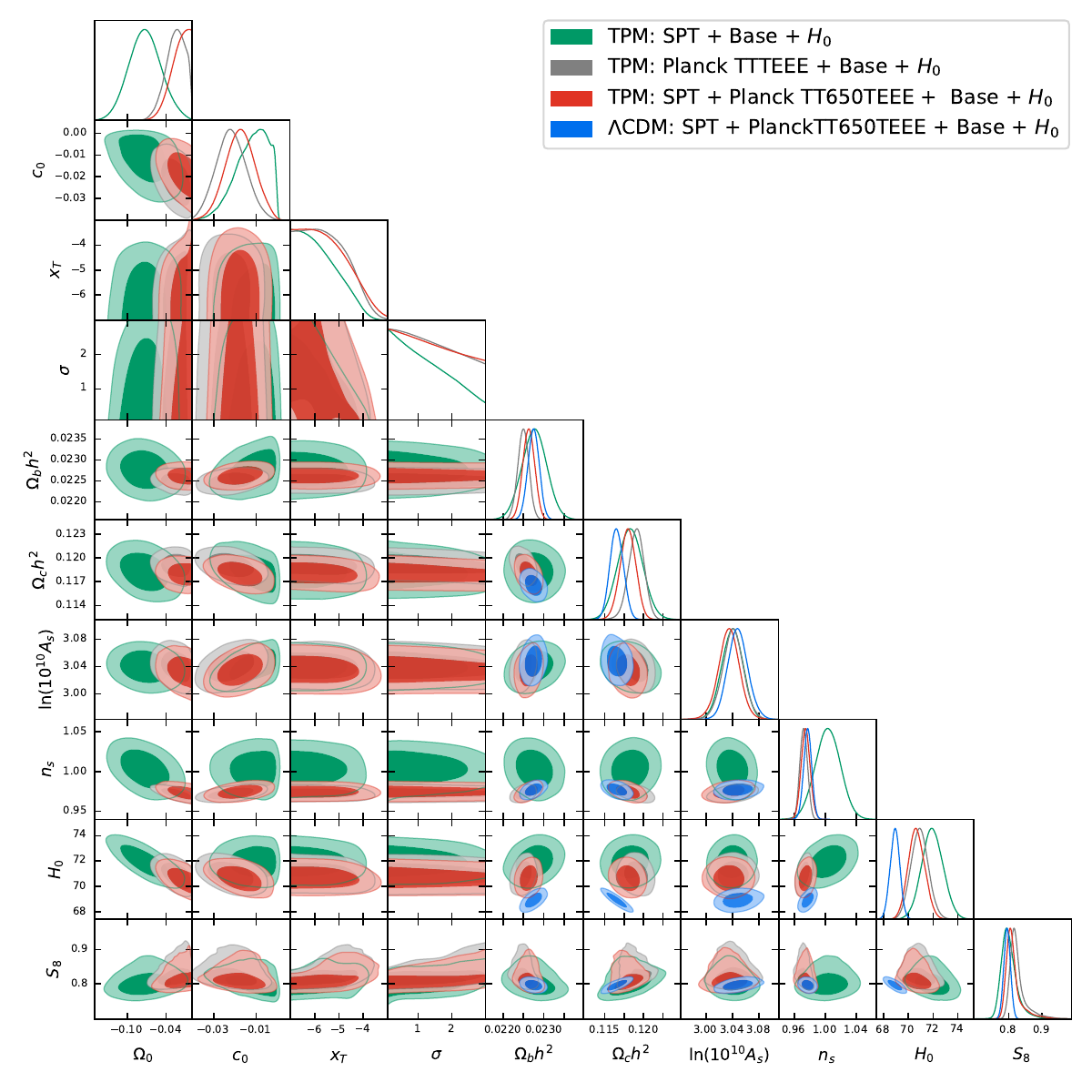}

\vspace{-3mm}

\caption{ The combination of \SPT\ + \TTCut\ + \base\ + \Hprior\ (red) constrains the TPM model more tightly than either the \TTTEEE\ + \base\ + \Hprior\ (gray) or \SPT\ + \base\ + \Hprior\ (green) data. While the \SPT\ + \base\ + \Hprior\ data prefer roughly $7\%$ transitions in value of the Planck mass, the inclusion of \TTCut\ restrict the amplitude of the transitions to be less than $5.1\%$ at the 95$\%$ confidence level. Additionally, it also restricts the allowed parameter space of $\Omega_bh^2$, $\Omega_ch^2$, and $n_s$. While the inclusion of \TTCut\ constrains the TPM model preferred values of $H_0$ to be lower than with SPT alone, the data still prefer $H_0 = 70.65 \pm 0.66$ \kmsmpc\ and $S_8 = 0.815 \pm 0.026$. Because the inclusion of \TTCut\ constrains $\Omega_0$ more tightly, the $c_0$ parameter, which acts as the dark energy equation of state in the late universe, decreases to allow for a higher $H_0$ value to fit the $H_0$ prior. The preferred value of $S_8$ for the TPM fit to \SPT\ + \TTCut\ + \base\ + \Hprior\ is slightly higher than the corresponding \lcdm\ preferred value of $S_8 = 0.7956 \pm 0.0099$ suggesting that the TPM model is not doing better than \lcdm\ (blue) on the clustering tension for these data. We cut the parameter values in the plot for $\Omega_0 > 0.0035$ for visualization purposes.
}

\label{fig:Comparison_SPT_PlikTT650TEEE}
\end{figure*}

In this subsection, we explore the impact of combining SPT and \planck\ anisotropy data. In particular, we run MCMCs for both \lcdm\ and TPM fits to the \SPT\ + \TTCut\ + \base\ + \Hprior\ likelihood combination (see Section~\ref{Data} for descriptions of each of these data sets). We show the resulting parameter constraints from the MCMCs in Table~\ref{tab:SPT_plik} and Figure~\ref{fig:Comparison_SPT_PlikTT650TEEE}. In Figure~\ref{fig:Comparison_SPT_PlikTT650TEEE}, we additionally include posteriors from the TPM fits to \SPT\ + \base\ + \Hprior, the TPM fit to \TTTEEE\ + \base\ + \Hprior, and the \lcdm\ fit to \SPT\ + \TTCut\ + \base\ + \Hprior\ for comparison. 

\begin{table*}[!tbp]
\setlength{\tabcolsep}{5pt}
\centering
\hspace{0cm}
\begin{tabular}{@{}cccccc@{}}

\toprule
 & $\Lambda$CDM: \SPT\  & $TPM$: \SPT\            \\
 & + \TTCut\  & + \TTCut\ \\
  & + \base\ + \Hprior\ & + \base\ + \Hprior\ \\
\toprule                                                                                                               
$100\theta_{\rm MC}$   & 1.04093 (1.04092 $\pm$ 0.00033)    & 1.04091 ($1.04098^{+0.00038}_{-0.00039}$)  \\
$\Omega_bh^2$          & 0.02275 (0.02275 $\pm$ 0.00014)    & 0.02261 (0.02263 $\pm$ 0.00014)   \\
$\Omega_ch^2$          & 0.11652 ($0.11655^{+0.00086}_{-0.00085}$)   &  0.1182 (0.1179 $\pm$ 0.0010)     \\
$\tau$                 & 0.0625 (0.0598 $\pm$ 0.0074)   &  0.0485 (0.0516 $\pm$ 0.0075)        \\
$\ln(10^{10}A_s)$      & 3.053 (3.048 $\pm$ 0.015)   &   3.027 (3.034 $\pm$ 0.015)       \\
$n_s$                  & 0.9771 (0.9771 $\pm$ 0.0050)   & 0.9733 (0.9746 $\pm$ 0.0057)     \\
\midrule

$\Omega_0$                  & -   &  -0.013 (> -0.051 at 95$\%$)       \\
$x_T$         & -   &  -5.57 ($-5.56^{+1.02}_{-1.01}$)          \\
$\sigma$         & -   & 1.42 ($1.44^{+1.00}_{-0.95}$)       \\

$c_0$              & -  & -0.0205 (> -0.0311 at 95$\%$)           \\
\midrule
$H_0$                  & 68.91 ($68.90^{+0.38}_{-0.39}$)    & 70.60 (70.65 $\pm$ 0.66)       \\
$\sigma_8$                  & 0.8044 (0.8026 $\pm$ 0.0064)  & 0.825 ($0.839^{+0.018}_{-0.020}$)      \\
$S_8$                  & 0.7971 (0.7956 $\pm$ 0.0099)   & 0.802 ($0.815^{+0.018}_{-0.020}$)     \\
$\chi^2_{\rm Planck TT650TEEE}$ & 445.73   & 445.53                  \\
$\chi^2_{\rm Planck low TT}$ & 21.48   & 21.31                  \\
$\chi^2_{\rm Planck lowE}$ & 397.74   & 395.73                  \\
$\chi^2_{\rm SPT}$ & 1122.85   & 1123.46                     \\
$\chi^2_{\rm CMB \ lensing}$ & 10.36   & 8.37                \\
$\chi^2_{\rm BAO}$ &  6.37  & 8.43                  \\
$\chi^2_{\rm Pantheon}$ & 1034.77   & 1037.11                      \\
$\chi^2_{\rm H_0}$ & 17.24   & 5.09                     \\
$\chi^2_{\rm prior}$ & 0.20   & 0.26                     \\
$\chi^2_{\rm tot}$ & 3056.76   & 3045.28                       \\
\end{tabular}
\caption{  Maximum Likelihood Parameter Values and, in Parenthesis, Mean plus 68$\%$ Confidence Level Bounds. Each Column Corresponds to the Model Fit to \SPT\ + \TTCut\ + \base\ + \Hprior\ Likelihood (See Section~\ref{Data} for Details of the Likelihoods).
} \label{tab:SPT_plik}
\end{table*}

The TPM model fits the \SPT\ + \TTCut\ + \base\ + \Hprior\ data better than \lcdm\ does with a combined $\Delta \chi^2 = -11.48$. This is a smaller improvement for the TPM model over \lcdm\ than the corresponding improvement in the fits to \SPT\ + \base\ + \Hprior\ as well as the fits to \TTTEEE\ + \base\ + \Hprior. Similar to both or these fits, most of the improvement found by the TPM model over \lcdm\ comes from a better fit to the $H_0$ prior. In this case, the improvement is $\Delta \chi^2 = -12.15$ meaning the TPM model fits the remaining data sets, collectively, slightly worse than \lcdm. We explore the effect of not including the $H_0$ prior in Appendix~\ref{SPT_H0} where we find the \planck\ anisotropy data remove the non-Gaussian tails in $H_0$ and $\Omega_0$ that allow for a resolution of the Hubble tension. We find that the best-fit value of $\Omega_0 = -0.0004$, which is consistent with no transition in the Planck mass. 

In general, the \SPT\ + \TTCut\ + \base\ + \Hprior\ restrict the TPM model to refind the preferred values from the TPM fit to \TTTEEE\ + \base\ + \Hprior. In particular, the constraint on $H_0 = 70.65 \pm 0.66$ \kmsmpc, which is very close to the constraint from \TTTEEE\ + \base\ + \Hprior\ given by $H_0 = 70.90^{+0.69}_{-0.70}$ \kmsmpc. Similarly, the constraints on $S_8$ are $S_8 = 0.815 \pm 0.026$ and $S_8 = 0.828 \pm 0.028$ for with and without SPT data respectively. The best-fit value of $S_8$ for the \lcdm\ fit to \SPT\ + \TTCut\ + \base\ + \Hprior\ is lower than the TPM best-fit value. 

While the best-fit parameters for the TPM fits to \TTTEEE\ + \base\ + \Hprior\ and \SPT\ + \TTCut\ + \base\ + \Hprior\ tend to be similar, the best-fit values of $\tau$ and ln($10^{10}A_s$) for the TPM fit to \SPT\ + \TTCut\ + \base\ + \Hprior\ are lower.

These shifts in the preferred values of $H_0$ and $S_8$ result from a tightening of parameter space for $\Omega_bh^2$, $\Omega_ch^2$, $n_s$, and $100\theta_{MC}$, which have comparable uncertainties to the uncertainties from the TPM fit to \TTTEEE\ + \base\ + \Hprior. The inclusion of \TTCut\ lowers the mean value of the scalar index to $n_s = 0.9746 \pm 0.0057$ compared to $n_s = 1.003 \pm 0.016$. Notably, the best-fit value of $n_s$ from the \lcdm\ fit to \SPT\ + \TTCut\ + \base\ + \Hprior\ is higher than the best-fit value for the TPM fit to the same data. 

For the TPM model specific parameters, the amplitude of the transition of the Planck mass, quantified by the $\Omega_0$ parameter, is constrained to be smaller than a $5.1\%$ shift at the 95$\%$ confidence level, which is a tighter constraint than when no SPT anisotropy data are included. Because the \SPT\ + \TTCut\ + \base\ + \Hprior\ constrain the amplitude of the transition in the Planck mass more tightly than \SPT\ + \base\ + \Hprior\ do, the $c_0$ parameter shifts to more negative values to fit the $H_0$ prior similar to the constraint on $c_0$ from the TPM fit to \planck\ TTTEEE +  Base + $H_0$. 

We show the results for the case where $c_0 = 0$, which we refer to as the TPM f(R) case, in Appendix~\ref{SPT_fR}. To summarize, we find that the amplitude of the transition of the Planck mass is constrained to be less than $6.7\%$ at the $95\%$ confidence level. This follows because the $H_0$ prior still shifts the preferred parameter space to find higher values of $H_0$, but the TPM f(R) case cannot modify the $c_0$ parameter. The constraint on $H_0 = 70.12 \pm 0.67$ \kmsmpc\ is roughly comparable to the full TPM model where the $c_0$ parameter is allowed to vary. 

The phenomenology of the transition quantified by the timing of the transition, $x_T$, and the rapidity of the transition $\sigma$, are still unconstrained for the TPM fit to SPT + \planck\ TT650TEEE + Base + $H_0$. 

We additionally tested the impact of combining SPT anisotropy data with the \TTTEEE\ likelihood. We find that we obtain qualitatively similar results with tighter constraints on parameters to the case when we use the \TTCut\ likelihood. The choice of fitting the TPM model to  \SPT\ + \TTCut\ + \base\ + \Hprior\ in this work was done to highlight the difference between TPM and EDE. For EDE, \SPT\ + \TTCut\ data have been shown to prefer higher values of $H_0$ even without including an $H_0$ prior from local measurements \citep{Smith/etal:2022}. This is not the case for the TPM model, which emphasizes the importance of testing each cosmological model with all of the available data sets to evaluate the viability of a cosmological model. 

Overall, we find that \textit{while SPT anisotropy data alone were less constraining of the TPM model and allowed for larger shifts in both the Planck mass and $H_0$, the inclusion of \TTCut\ restricts the allowed parameter space to be similar to the TPM fit to \TTTEEE\ + \base\ + \Hprior\ limit, though with tighter constraints on the amplitude of the transition of the Planck mass. Moreover, the $\chi^2$ improvement found by the TPM model over \lcdm\ for the combination of \planck\ and SPT data is smaller than for either data set alone. }

\subsection{Understanding the Differences Between Preferred Cosmology by SPT and \Planck} \label{SPT_discussion}

In this subsection, we explore the differences between the preferred parameter values for the TPM fit to \SPT\ + \base\ + \Hprior\ and the TPM fit to \TTTEEE\ + \base\ + \Hprior. In particular, we run an additional MCMC for the TPM fit to \TTTEEE\ + \base\ + \Hprior\ where we allow the helium fraction, $Y_{He}$, to vary. As discussed in \cite{Hou/etal:2013}, the helium fraction only enters the CMB anisotropy calculations through the ionization fraction of electrons, which directly affects the CMB damping tail. While the helium fraction does have some effect on the size of sound horizon, this is subdominant to the effect on the damping scale. 

\begin{figure*}[!tbp]
\centering

\hspace{-1cm}
\includegraphics[clip,width=\linewidth]{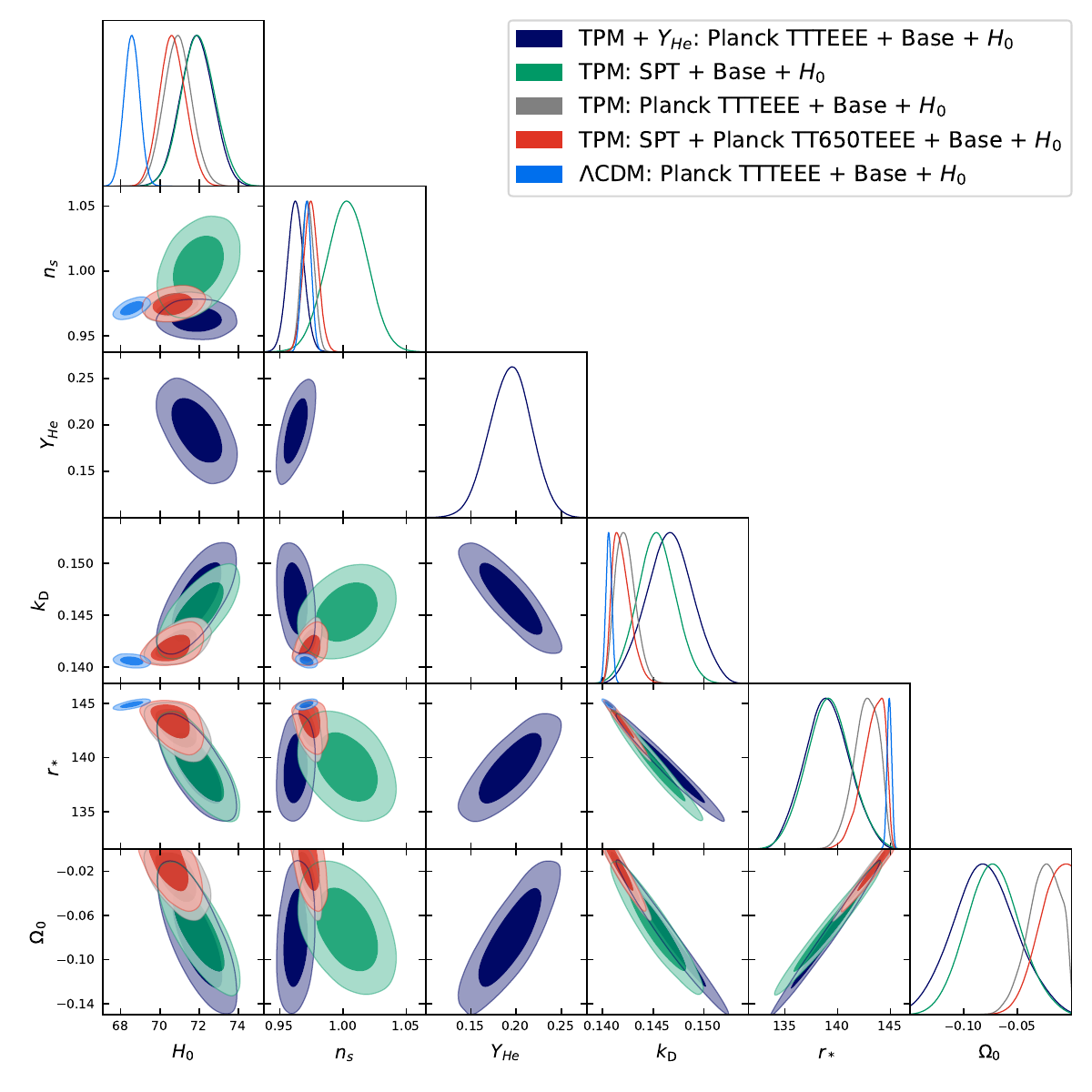}

\vspace{-3mm}

\caption{ Allowing the helium fraction to vary within the TPM model when fit to \TTTEEE\ + \base\ + \Hprior\ (dark blue) opens up the degeneracy between the sound horizon, $r_*$, and $H_0$ to match the degeneracy found when the TPM model is fit to \SPT\ + \base\ + \Hprior\ (green). Note that the preferred values of the helium fraction found by this fit are not physical, but this test illuminates the differences between fitting to SPT and \planck\ data. The helium fraction only enters the CMB calculations in the ionization fraction of electrons and primarily affects the photon damping scale, $k_D$. Larger amplitude transitions in the TPM model lead to suppressed power on small angular scales. Both \planck\ and SPT small angular scale anisotropy data disfavor this suppression in power meaning other parameters have to shift to compensate. For the TPM model fit to \SPT\ + \base\ + \Hprior, this can be compensated for by increasing $n_s$ to have more primordial power on small scales. However when the TPM model is fit to \TTTEEE\ + \base\ + \Hprior\ (gray) or \SPT\ + \TTCut\ + \base\ + \Hprior\ (red), $n_s$ is restricted by the lower multipole data in combination with the higher multipole data, which prevents the model from finding higher $H_0$ and lower $r_*$ values. We cut the parameter values in the plot for $\Omega_0 > 0.0035$ for visualization purposes.
}

\label{fig:TPM_Yhe}
\end{figure*}

We show the results of the TPM + $Y_{He}$ fit to \TTTEEE\ + \base\ + \Hprior\ in Table~\ref{tab:Y_he} and Figure~\ref{fig:TPM_Yhe}. The constraint on $Y_{He}$ for the TPM + $Y_{He}$ fit to \TTTEEE\ + \base\ + \Hprior\ is given by $Y_{He} = 0.195 \pm 0.023$, which is $2.3 \sigma$ lower than the BBN consistency relation between $Y_{He}$ and the physical baryon density even given the larger uncertainties when the helium fraction is allowed to vary. Additionally, these low values of $Y_{He}$ conflict with more direct empirical estimates of the primordial helium abundance \cite[see, e.g.][for a review]{fields/etal:2020}. The TPM fits with $Y_{He}$ varying are nevertheless valuable for elucidating the differences between the constraining power of SPT and \planck\ data. 

Allowing the helium fraction to vary in the TPM fit to \TTTEEE\ + \base\ + \Hprior\ results in a further opening of the degeneracy between the sound horizon, $r_*$, and $H_0$ to match the degeneracy found when TPM is fit to \SPT\ + \base\ + \Hprior\ as shown in Figure~\ref{fig:TPM_Yhe}. Because the primary effect on CMB anisotropy calculations of varying the helium mass is to affect the physics of diffusion damping on small scales, and decreasing the helium fraction results in an increase in power on small scales or less diffusion damping, we conclude that the TPM model prefers a suppression of power on small angular scales that is disfavored by \TTTEEE\ data. Varying the helium fraction within the TPM model means there are now two new parameters, $\Omega_0$ and $Y_{He}$, that affect the small angular scale CMB power spectra in ways that can cancel out.  

This suppression of power on small angular scales by the TPM model is consistent with results shown in Figure 3 of B22, which shows that the best-fit TPM model prefers a suppression of power on small scales in both the TT and EE power spectra. This occurs because the angular size of the sound horizon, $\theta_* = r_*/D_A$, and the angular size of the damping scale, $\theta_D = r_D/D_A$, are both dependent on the same angular diameter distance to the surface of last scattering. The CMB measurements constrain the angles $\theta_*$ and $\theta_D$ meaning lowering the sound horizon $r_*$ results in a drop of $D_A$ to compensate. This in turn shifts the physical size of the damping scale, $r_D$, which sets the physical scale of CMB anisotropy suppression.

So do SPT data prefer this suppression of power on small angular scales? Fitting the TPM model to \SPT\ + \base\ + \Hprior\ data results in a large increase in $n_s$. This suggests that the SPT data also do not favor the suppression of power on small angular scales favored by larger amplitude transitions in the Planck mass, but that the data allow for shifts in parameters like $n_s$. Increasing $n_s$ shifts power in the primordial power spectrum from large angular scales to small angular scales meaning that more negative transitions in the Planck mass, which are necessary to resolve the Hubble tension within the TPM model, can be compensated for by increasing $n_s$. This can be seen in the increases in the preferred value of $n_s$ for the TPM model compared to \lcdm\ for both fits to \SPT\ + \base\ + \Hprior\  and \TTTEEE\ + \base\ + \Hprior, which shown in Table~\ref{tab:parameters_SPT_no_plik}. 

When the TPM model is fit to the \SPT\ + \base\ + \Hprior\ data,
$n_s$ is allowed to increase so that $n_s > 1$, which in turn allows for large enough shifts in the Planck mass to bring $H_0$ into agreement with the $H_0$ prior. This results from the fact that SPT anisotropy data do not have a low multipole anchor to constrain $n_s$ at the same level as the full \planck\ anisotropy data.

Therefore, there is more freedom in parameter space to increase $n_s$ and shift power from large angular scales to small angular scales. However, additionally including \TTCut\ constrains the large angular scales and complements the SPT data allowing for a tighter constraint on $n_s$. This is why the best-fit values of $n_s$ and as a result $H_0$ are lower when \TTCut\ are added to SPT data. 

\textit{In both the TPM fit to \SPT\ + \base\ + \Hprior\ data and the \TTTEEE\ + \base\ + \Hprior\ data, the CMB anisotropy data do not prefer the suppression of power on small angular scales necessary to resolve the Hubble tension within the TPM model, but SPT data allow for larger shifts of parameters like $n_s$, which allows it to compensate for the small-scale suppression of power.}

\begin{table*}[!tbp] 
\setlength{\tabcolsep}{5pt}
\centering
\hspace{0cm}
\begin{tabular}{@{}cccccc@{}}

\toprule
& $\Lambda$CDM                    &  $\Lambda$CDM + $Y_{He}$                 &   $TPM$ & $TPM$ + $Y_{He}$         \\
\toprule                                                                                                               
$100\theta_{\rm MC}$   & 1.0413 ($1.04127^{+0.00029}_{-0.00030}$ )   &  1.0416 ($1.0416^{+0.00048}_{-0.00049}$)  & 1.4139 (1.04135 $\pm$ 0.00036 ) &  1.04064 (1.04038 $\pm$ 0.00055 )  \\
$\Omega_bh^2$          & 0.002261 (0.02261 $\pm$ 0.00013)   &  0.022699 (0.022692 $\pm$ 0.00017)  & 0.022505 (0.022498 $\pm$ 0.00013) &  0.02219 (0.02210 $\pm$ 0.00021)  \\
$\Omega_ch^2$          & 0.11748 (0.11741 $\pm$ 0.00086)   &  0.1177 (0.1176 $\pm$ 0.00086)  & 0.1193 (0.11906 $\pm$ 0.00099) &  0.1190 ($0.1192^{+0.00099}_{-0.00098}$)     \\
$\tau$                 & 0.0615 ($0.0633^{+0.0080}_{-0.0079}$)   & 0.0631 ($0.0639^{+0.0080}_{-0.0078}$)  & 0.0532 (0.0528 $\pm$ 0.0074) & 0.0525 ($0.0520^{+0.0075}_{-0.0076}$)        \\
$\ln(10^{10}A_s)$      & 3.054 ($3.058^{+0.016}_{-0.015}$)   & 3.059 (3.062 $\pm$ 0.016)  & 3.043 (3.040 $\pm$ 0.015) & 3.035 (3.034 $\pm$ 0.016)        \\
$n_s$                  & 0.9712 (0.9713 $\pm$ 0.0036)   & 0.9753 ($0.9754^{+0.0060}_{-0.0059}$)  & 0.9715 (0.9721 $\pm$ 0.0048) &  0.9654 (0.9627 $\pm$ 0.0063)     \\
\midrule

$\Omega_0$                  & -   & -  & -0.025 (> -0.058 at 95$\%$) &  -0.069 (-0.080 $\pm$ 0.029)      \\
$x_T$         & -   & -  & -5.33 (-5.58 $\pm$ 0.99) &  -4.61 ($-5.81^{+0.87}_{-0.83}$)         \\
$\sigma$         & -   & -  & 0.82 ($1.42^{+0.98}_{-0.93}$) & 0.924 ($1.32^{+0.96}_{-0.88}$)        \\

$c_0$              & -   & - & -0.02293 (-0.02174 $\pm$ 0.0071) & -0.0183 (>-0.031 at 95$\%$)           \\
\midrule
$H_0$                  & 68.56 (68.57 $\pm$ 0.39)   & 68.69 (68.71 $\pm$ 0.42 )   & 70.94 ($70.90^{+0.69}_{-0.70}$) & 71.74 (71.87 $\pm$ 0.84)       \\
$\sigma_8$                  & 0.8076 (0.8079 $\pm$ 0.0064)   & 0.8115 ($0.8123^{+0.0074}_{-0.0073}$) & 0.839 ($0.853^{+0.020}_{-0.021}$) &  0.853 ($0.841^{+0.020}_{-0.021}$)         \\
$S_8$                  & 0.8068 ($0.8072^{+0.0098}_{-0.0099}$)  & 0.810 (0.810 $\pm$ 0.010)  & 0.815 ($0.828^{+0.020}_{-0.021}$) & 0.818 ($0.805^{+0.022}_{-0.023}$)        \\
$Y_{He}$                  & 0.246810 ($0.246810^{+0.000047}_{-0.000048}$)  & 0.257 (0.257 $\pm$ 0.012)  & 0.246761 (0.246770 $\pm$ 0.000051) &  0.206 (0.195 $\pm$ 0.023)        \\
$\chi^2_{\rm Planck TTTEEE}$ & 589.78   & 590.63 & 585.69 & 585.22                 \\
$\chi^2_{\rm Planck low TT}$ & 22.41   & 21.92  & 21.41 & 21.29                \\
$\chi^2_{\rm Planck lowE}$ & 397.70   & 398.24  & 395.83 & 395.83                  \\
$\chi^2_{\rm CMB \ lensing}$ & 9.575   & 9.46  & 8.82 & 8.27                \\
$\chi^2_{\rm BAO}$ & 5.55   & 5.65  & 7.74 & 6.7                    \\
$\chi^2_{\rm Pantheon}$ & 1034.73   & 1034.73  & 1037.08 & 1035.89                     \\
$\chi^2_{\rm H_0}$ & 20.68   & 19.36  & 3.52 & 0.96                   \\
$\chi^2_{\rm prior}$ & 0.22   & 0.04  & 0.09 & 0.03                   \\
$\chi^2_{\rm tot}$ & 2081.39   & 2080.03  & 2060.19 & 2054.19                    \\
\end{tabular}
\caption{ Maximum Likelihood Parameter Values and, in Parenthesis, Mean plus 68$\%$ Confidence Level Bounds. Each Column Corresponds to the Model Fit to \TTTEEE\ + \base\ + \Hprior\ Likelihood (See Section~\ref{Data} for Details of the Likelihoods).
} \label{tab:Y_he}
\end{table*}

\section{Including Additional Large Scale Structure Data: BOSS Redshift Space Distortions and DES Y1 Cosmic Shear, Galaxy-Galaxy Lensing, and Galaxy Clustering} \label{Results_LSS}

In this section, we explore how additional LSS data can be used to constrain the TPM model. In particular, we look at two additional data sets, the BOSS \RSD\ and \DES\ cosmic shear, galaxy-galaxy lensing, and galaxy clustering measurements. We show the results of including \RSD\ measurements in Section~\ref{RSD} and the results of including the \DES\ cosmic shear, galaxy-galaxy lensing, and galaxy clustering measurements in Section~\ref{DES}.

\subsection{BOSS Redshift Space Distortions} \label{RSD}

\begin{figure*}[!tbp]
\centering

\hspace{-1cm}
\includegraphics[clip,width=\linewidth]{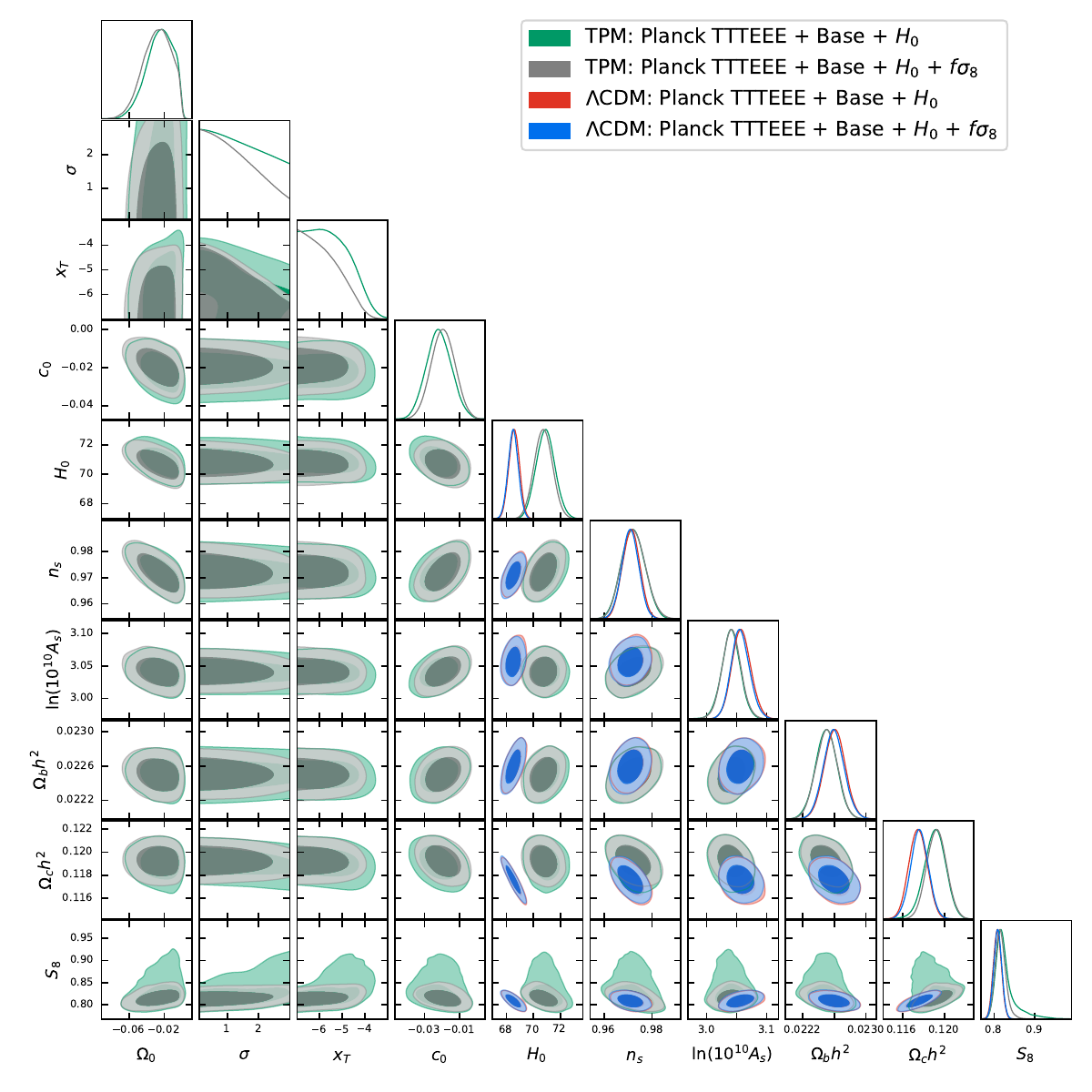}

\vspace{-3mm}

\caption{ Including RSD $f\sigma_8$ data results in minimal changes to the posterior parameter distribution functions, but including RSD data does remove the non-Gaussian tail present in the parameter posterior for $S_8$ for the TPM model fit \TTTEEE\ + \base\ + \Hprior. In Figure 5 of B22, we showed that the part of allowed parameter space where $S_8 > 0.85$ correspond to later (closer to recombination) transitions in the Planck mass. This is supported by the slight restriction of allowed parameter space for $x_T$ to disfavor $x_T > -4$ in the TPM fit to \TTTEEE\ + \base\ + \RSD\ + \Hprior\ relative to the case without the \RSD\ likelihood. The RSD data still allow for parameter value of $H_0 > 70$ \kmsmpc\ and $S_8 < 0.8$, which would alleviate both the Hubble and clustering tensions. 
}

\label{fig:fsig8}
\end{figure*}

\begin{table*}[!tbp]
\setlength{\tabcolsep}{5pt}
\centering
\hspace{-1cm}
\begin{tabular}{@{}cccccc@{}}

\toprule

 & $\Lambda$CDM: \TTTEEE\                        & $TPM$: \TTTEEE\                      & $\Lambda$CDM: \TTTEEE\   & $TPM$: \TTTEEE\  \\ 
& + \base\ + \Hprior\ & + \base\ + \Hprior\ & + \base\ + \Hprior\ + \texttt{RSD} & + \base + \Hprior\ + \texttt{RSD} \\
\toprule                                                                                                               
$100\theta_{\rm MC}$   & 1.04130 ($1.04127^{+0.00029}_{-0.00030}$)   & 1.04139 (1.04135 $\pm$ 0.00036)  & 1.04121 (1.04124 $\pm$ 0.00029) & 1.04141 (1.04138 $\pm$ 0.00036) \\
$\Omega_bh^2$          & 0.02261 (0.02261 $\pm$ 0.00013)   & 0.022505 (0.022498 $\pm$ 0.00013)  & 0.02256 (0.02259 $\pm$ 0.00013) & 0.022514 (0.022503 $\pm$ 0.00013)  \\
$\Omega_ch^2$          & 0.11748 (0.11748 $\pm$ 0.00086)   & 0.1193 (0.11906 $\pm$ 0.00099)  & 0.1178 ($0.1176^{+0.00083}_{-0.00082}$) & 0.11913 (0.11924 $\pm$ 0.00092)    \\
$\tau$                 & 0.0615 ($0.0633^{+0.0080}_{-0.0079}$)   & 0.0532 (0.0528 $\pm$ 0.0074 )  & 0.0601 ($0.0624^{+0.0078}_{-0.0077}$) & 0.0543 (0.0530 $\pm$ 0.0076)        \\
$\ln(10^{10}A_s)$      & 3.054 ($3.058^{+0.016}_{-0.015}$)   & 3.043 (3.040 $\pm$ 0.015)  & 3.053 (3.056 $\pm$ 0.015) & 3.044 (3.041 $\pm$ 0.015)        \\
$n_s$                  & 0.9712 (0.9713 $\pm$ 0.0036)   & 0.9715 (0.9721 $\pm$ 0.0048)  & 0.9700 (0.9709 $\pm$ 0.0035) & 0.9721 (0.9723 $\pm$ 0.0049)     \\
\midrule
$\Omega_0$                  & -   & -0.025 (> -0.058 at 95$\%$)  & - & -0.028 (> -0.061 at 95$\%$)       \\
$x_T$         & -   & -5.33 (-5.58 $\pm$ 0.99)  & - & -5.58 (-$5.84^{+0.87}_{-0.85}$)         \\
$\sigma$         & -   & 0.82 ($1.42^{+0.98}_{-0.93}$)  & - & 1.03 ($1.24^{+0.89}_{-0.83}$)       \\
$c_0$              & -   & -0.02293 (-0.02174 $\pm$ 0.0071) & - & -0.0196 (> -0.0323 at 95$\%$)           \\
\midrule
$H_0$                  & 68.56 (68.57 $\pm$ 0.39)   & 70.94 ($70.90^{+0.69}_{-0.70}$ )   & 68.40 (68.50 $\pm$ 0.38) & 70.81 (70.74 $\pm$ 0.68)       \\
$\sigma_8$                  & 0.8076 (0.8091 $\pm$ 0.0064)   & 0.839 ($0.853^{+0.020}_{-0.021}$) & 0.8077 (0.8088 $\pm$ 0.0062) & 0.834 ($0.838^{+0.010}_{-0.011}$)       \\
$S_8$                  & 0.8068 ($0.8083^{+0.0098}_{-0.0099}$)   & 0.815 ($0.828^{+0.020}_{-0.021}$)  & 0.8094 (0.8092 $\pm$ 0.0097) & 0.811 ($0.816^{+0.012}_{-0.013}$)       \\
$\chi^2_{\rm Planck TTTEEE}$ & 589.78   & 585.69 & 588.73 & 586.83                 \\
$\chi^2_{\rm Planck low TT}$ & 22.41   & 21.41  & 22.58 & 21.30                \\
$\chi^2_{\rm Planck lowE}$ & 397.70   & 395.83  & 397.30 & 396.00                  \\
$\chi^2_{\rm CMB \ lensing}$ & 9.575   & 8.82  & 9.45 & 8.46               \\
$\chi^2_{\rm BAO}$ & 5.55   & 7.74  & 7.94 & 11.49                   \\
$\chi^2_{\rm Pantheon}$ & 1034.73   & 1037.08  & 1034.75 & 1036.39                     \\
$\chi^2_{\rm H_0}$ & 20.68   & 3.52  & 22.36 & 4.10                   \\
$\chi^2_{\rm prior}$ & 0.22   & 0.09  & 0.31 & 0.00                   \\
$\chi^2_{\rm tot}$ & 2081.39   & 2060.19  & 2083.42 & 2064.56                     \\
\end{tabular}
\caption{ \label{tab:fsig8} Maximum Likelihood Parameter Values and, in Parenthesis, Mean plus 68$\%$ Confidence Level Bounds. Each Column Delineates the Model and Likelihood Combination (See Section~\ref{Data} for Details of the Likelihoods).
}
\end{table*}

We use the $f\sigma_8$ constraints from BOSS DR 12 measurements in \cite{alam/etal:2017} to constrain the TPM model. In particular, we run an MCMC of the TPM model fitting the \TTTEEE\ + \base\ + \RSD\ + \Hprior\ (see Section~\ref{Data} for details of the \RSD\ likelihood). The BOSS DR 12 RSD measurements primarily constrain the parameter combination $f\sigma_8$ at redshifts 0.38, 0.51, and 0.61. Here $f \equiv dln(D)/dln(a)$ is the linear growth rate where $D$ is the growth factor related to the density perturbations $\delta$. We show the results of this MCMC in Figure~\ref{fig:fsig8} and Table~\ref{tab:fsig8}.

In general, we find that the RSD measurements do not add significant new constraints to the TPM model over the existing TPM constraints from \TTTEEE\ + \base\ + \Hprior. The primary change is that the non-Gaussian tail present in the posterior distribution function for $S_8$ for the TPM model fit to the \TTTEEE\ + \base\ + \Hprior\ is removed. 

In B22, we showed that $S_8$ values in this non-Gaussian tail result from transitions that happen closer to recombination. This is supported by Figure~\ref{fig:fsig8}, which shows that the values of $x_T$ that are closest to recombination are removed from the posteriors for the TPM fit to \TTTEEE\ + \base\ + \Hprior\ + \RSD\ compared to the posteriors for the TPM fit to \TTTEEE\ + \base\ + \Hprior. 

\begin{figure}[!tbp]
\centering

\hspace{1cm}
\includegraphics[clip,width=\linewidth]{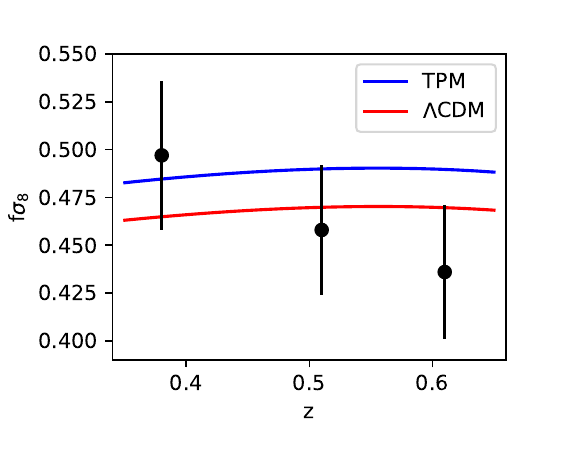}

\vspace{-3mm}

\caption{ The best-fit TPM model fit to \TTTEEE\ + \base\ + \RSD\ + \Hprior\ (blue) prefers an enhanced growth of structure compared to the best-fit \LCDM\ model (red) to the same data. The data points (black) are BOSS DR 12 f$\sigma_8$ constraints at redshifts 0.38, 0.51, and 0.61 \citep{alam/etal:2017}. In the TPM model, later transitions in the Planck mass result in longer wavelength modes experiencing enhanced gravity, which increases the amount of clustering on these scales. The RSD data rule out the late transitions that lead to values of $S_8 \gtrapprox 0.85$ (see Figure~\ref{fig:fsig8}) because these transitions enhance the growth of structure to a point that is strongly disfavored. 
}

\label{fig:fsig8_vs_z}
\end{figure}

In the TPM model, matter perturbation modes that enter the horizon during the transition in the Planck mass are enhanced relative to those modes that enter after the transition. Therefore, later transitions (closer to recombination) result in enhanced matter perturbation modes for larger wavelength or smaller wavenumber modes relative to the \lcdm\ case. This in turn affects the growth of structure and in general leads to an increased $f\sigma_8$ as shown in Figure~\ref{fig:fsig8_vs_z}.

The TPM model fits the \TTTEEE\ + \base\ + \RSD\ + \Hprior\ better than \lcdm\ with a cumulative $\Delta \chi^2 = 18.86$, which is smaller than the equivalent improvement when the RSD data are not included, $\Delta \chi^2 = 21.20$.

\textit{While the addition of the $f\sigma_8$ constraints from RSD measurements are not constraining enough to rule out the TPM model entirely, these RSD measurements do disfavor later transitions (closer to recombination) that result in relatively large values of $S_8$. As a result, the part of the TPM parameter space that allows for $H_0 > 70$ \kmsmpc\ and $S_8 < 0.80$ when fit to \TTTEEE\ + \base\ + \Hprior\ are still allowed.} 

\subsection{DES Y1} \label{DES}

\begin{table*}[!tbp]
\setlength{\tabcolsep}{5pt}
\centering
\hspace{-1cm}
\begin{tabular}{@{}cccccc@{}}

\toprule

 & $\Lambda$CDM: \TTTEEE\                        & $TPM$ f(R): \TTTEEE\                      & $\Lambda$CDM: \TTTEEE\   & $TPM$ f(R): \TTTEEE\  \\ 
& + \base\ + \Hprior\ & + \base\ + \Hprior\ & + \base\ + \Hprior\ + \texttt{DES} & + \base\ + \Hprior\ + \texttt{DES} \\
\toprule                                                                                                               
$100\theta_{\rm MC}$   & 1.04130 ($1.04127^{+0.00029}_{-0.00030}$)   & 1.04186 (1.04184 $\pm$ 0.00034)  &  1.04128 (1.04129 $\pm$ 0.00029) & 1.04178 (1.04185 $\pm$ 0.00034) \\
$\Omega_bh^2$          & 0.02261 (0.02261 $\pm$ 0.00013)   & 0.02264 (0.02262 $\pm$ 0.00013)  &  0.022635 (0.022642 $\pm$ 0.00013) & 0.022659 (0.022660 $\pm$ 0.00013)  \\
$\Omega_ch^2$          & 0.11748 (0.11748 $\pm$ 0.00086)   & 0.11832 ($0.11812^{+0.00091}_{-0.00090}$)  &  0.11717 ($0.11702^{+0.00083}_{-0.00081}$) & 0.11796 (0.11782 $\pm$ 0.00086)    \\
$\tau$                 & 0.0615 ($0.0633^{+0.0080}_{-0.0079}$)   &  0.0621 (0.0626 $\pm$ 0.0077) & 0.0617 ($0.0619^{+0.0077}_{-0.0075}$) & 0.0621 (0.0618 $\pm$ 0.0075)        \\
$\ln(10^{10}A_s)$      & 3.054 ($3.058^{+0.016}_{-0.015}$)   & 3.061 (3.062 $\pm$ 0.015)  &  3.053 (3.054 $\pm$ 0.015) & 3.062 (3.060 $\pm$ 0.015)        \\
$n_s$                  & 0.9712 (0.9713 $\pm$ 0.0036)   & 0.9791 (0.9800 $\pm$ 0.0045)  & 0.9717 ($0.9723^{+0.0035}_{-0.0036}$) & 0.9802 (0.9806 $\pm$ 0.0044)     \\
\midrule

$\Omega_0$                  & -   & -0.051 (-0.050 $\pm$ 0.016)  & - & -0.049 (-0.050 $\pm$ 0.015)       \\
$x_T$         & -   & -5.52 (-$5.50^{+0.95}_{-1.01}$)  & - & -5.66 (-5.66 $\pm$ 0.92)         \\
$\sigma$         & -   & 1 (fixed)  & - & 1 (fixed)     \\
$c_0$              & -   & 0 (fixed) & - & 0 (fixed)          \\
\midrule
$H_0$                  & 68.56 (68.57 $\pm$ 0.39)   & 70.28 ($70.33^{+0.68}_{-0.69}$)  &  68.70 (68.76 $\pm$ 0.37) & 70.35 (70.45 $\pm$ 0.65)       \\
$\sigma_8$                  & 0.8076 (0.8091 $\pm$ 0.0064)   & 0.816 ($0.828^{+0.018}_{-0.016}$) &  0.8064 (0.8064 $\pm$ 0.0061) & 0.816 ($0.821^{+0.010}_{-0.012}$)       \\
$S_8$                  & 0.8068 ($0.8083^{+0.0098}_{-0.0099}$)   & 0.798 ($0.808^{+0.019}_{-0.018}$)  &  0.8032 ($0.8020^{+0.0093}_{-0.0092}$) & 0.796 ($0.800^{+0.013}_{-0.014}$)       \\
$\chi^2_{\rm Planck TTTEEE}$ & 589.78   & 595.08 & 589.58 & 593.73                 \\
$\chi^2_{\rm Planck low TT}$ & 22.41   & 20.73  & 22.26 & 20.62                \\
$\chi^2_{\rm Planck lowE}$ & 397.70   & 398.00  & 397.71 & 397.95                  \\
$\chi^2_{\rm CMB \ lensing}$ & 9.575   & 9.28  & 9.28 & 9.16               \\
$\chi^2_{\rm BAO}$ & 5.55   & 5.28  & 5.82 & 5.44                   \\
$\chi^2_{\rm Pantheon}$ & 1034.73   & 1034.75  & 1034.74 & 1034.74                     \\
$\chi^2_{\rm H_0}$ & 20.68   & 6.83  & 19.28 & 6.45                   \\
$\chi^2_{\rm DES}$ & -   & -  & 320.28 & 320.66                   \\
$\chi^2_{\rm prior}$ & 0.22   & 0.14  & 1.59 & 1.79                   \\
$\chi^2_{\rm tot}$ & 2081.39   & 2070.09  & 2400.63 & 2390.54                     \\
\end{tabular}
\caption{ \label{tab:DES} Maximum Likelihood Parameter Values and, in Parenthesis, Mean plus 68$\%$ Confidence Level Bounds. Each Column Delineates the Model and Likelihood Combination (See Section~\ref{Data} for Details of the Likelihoods).
}
\end{table*}

\begin{figure*}[!tbp]
\centering

\hspace{-1cm}
\includegraphics[clip,width=\linewidth]{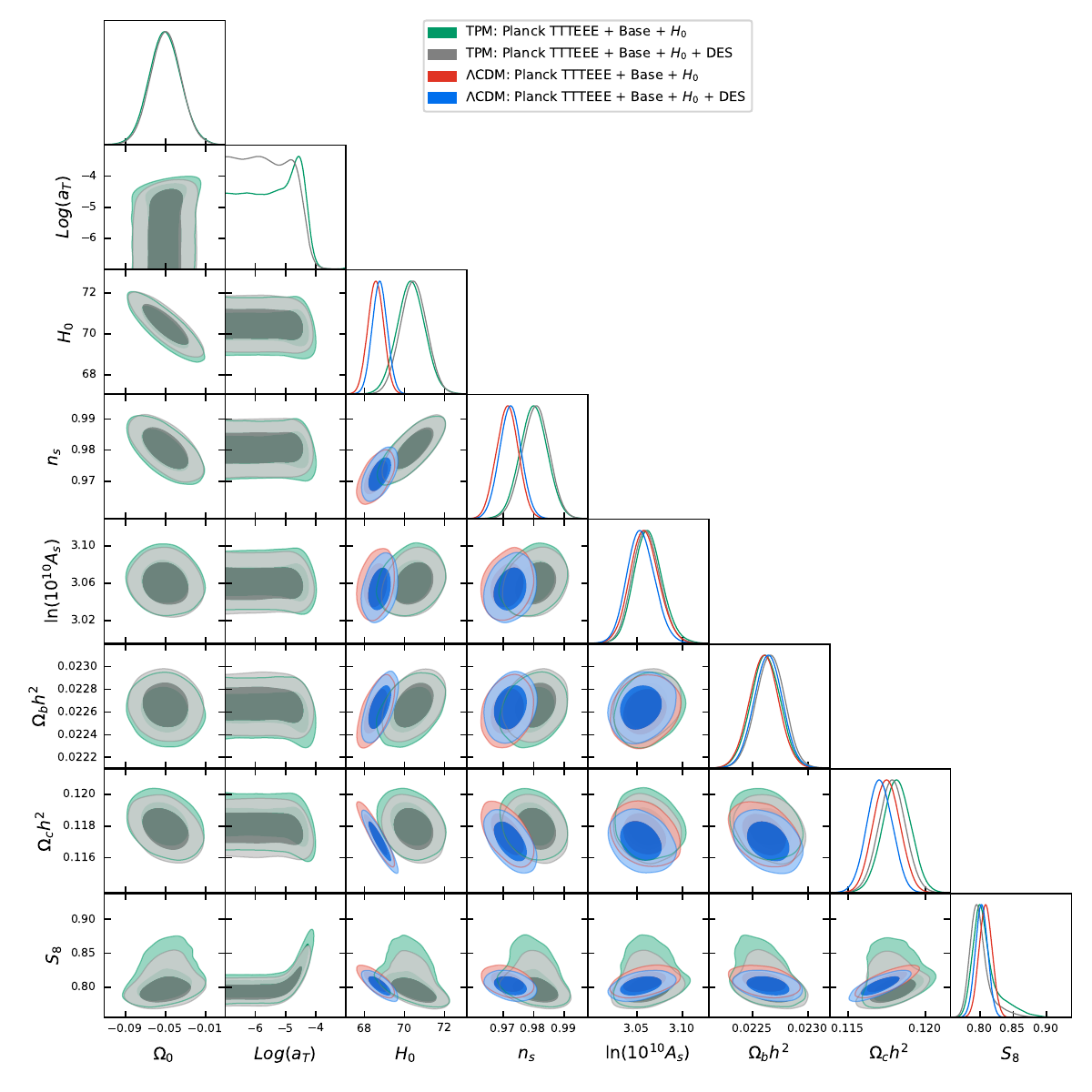}

\vspace{-3mm}

\caption{ We include DES Y1 cosmic shear, galaxy-galaxy lensing, and galaxy clustering measurements, collectively referred to as the 3x2 point analysis in fits of the TPM model. In all cases, we use scale cuts on DES data to remove non-linear scales as we do not have a TPM model theory for the non-linear collapse of dark matter. Additionally, we fix the TPM parameter $c_0 = 0$, which corresponds to the TPM f(R) case, as well as $\sigma = 1$ for numerical reasons. The inclusion of DES Y1 3x2 data do not significantly tighten the constraints on the TPM model parameters compared to the existing constraints from CMB, BAO, and Supernova data. Similar to the inclusion of RSD data, the DES Y1 3x2 data restricts some of the non-Gaussian tail for $S_8$. 
}

\label{fig:DES}
\end{figure*}

In this subsection, we explore how including weak lensing data can constrain the TPM model. In particular, we combine \DES\ cosmic shear, galaxy-galaxy lensing, and galaxy clustering data with \TTTEEE\ + \base\ + \Hprior. This collection of weak lensing data is typically referred to as a 3x2 point analysis. The Dark Energy Survey is a photometric survey. The \DES\ data include measurements from the first year of the survey and cover 1321 deg$^2$ of the sky. See Section~\ref{Data} for details of the \DES\ data including cuts in wavenumber that we make to only include linear scales. 

We note that while the TPM model tends to prefer lower values of $S_8$ when fit to \TTTEEE\ + \base\ + \Hprior\ data, it is unclear whether this will actually be in better agreement with weak lensing data. In particular, B22 showed that the TPM model lowers $S_8$ relative to \lcdm\ when fit to \TTTEEE\ + \base\ + \Hprior\ data by lowering the fractional matter density, $\Omega_m$, while $\sigma_8$ either does not change much or increases relative to \lcdm. Notably, in \cite{DES+KiDS}, it is shown that there is good agreement between cosmic shear measurements and \planck\ on the determination of $\sigma_8$ but not $S_8$ or $\Omega_m$ highlighting a possibility that the TPM model could fit both data better. It is therefore of interest to see how the TPM model compares when fit to actual cosmic shear data. 

We run an MCMC of the TPM model fit to \TTTEEE\ + \base\ + \DES\ + \Hprior\ data. For this MCMC, we restrict $c_0 = 0$, which corresponds to the TPM f(R) case. We additionally restrict $\sigma = 1$. These choices were made because we found numerical instabilities when evaluating the \DES\ likelihood. Sometimes numerical instabilities correspond to unphysical situations, but that is not the cause in this case. For only the TPM fit to \TTTEEE\ + \base\ + \DES\ + \Hprior, we use a Gelman-Rubin convergence criteria of R-1 = 0.02. For comparison, we also run an MCMC of the \lcdm\ model fit to \TTTEEE\ + \base\ + \DES\ + \Hprior\ data. We show the results of these MCMCs in Figure~\ref{fig:DES} and Table~\ref{tab:DES}.

In general, we find similar results for the TPM fit to \TTTEEE\ + \base\ + \DES\ + \Hprior\ as for the TPM fit to \TTTEEE\ + \base\ + \RSD\ + \Hprior. The primary effect of including the \DES\ data is to reduce the upper bounds for the $S_8$ parameter. In this case, the 95$\%$ confidence interval for the TPM fit to \TTTEEE\ + \base\ + \DES\ + \Hprior\ is higher than the 95$\%$ confidence interval for the TPM fit to \TTTEEE\ + \base\ + \RSD\ + \Hprior. The $1\sigma$ uncertainty for $S_8$ is approximately 1.4 times smaller for the TPM fit to \TTTEEE\ + \base\ + \DES\ + \Hprior\  compared to the TPM fit to \TTTEEE\ + \base\ + \Hprior. 

In summary \textit{we find that the \DES\ data are not sufficiently constraining to significantly affect the TPM model parameter space allowed by \planck, BAO, and Supernova measurements, though \DES\ data does slightly restrict the non-Gaussian tail for the $S_8$ parameter similar to the inclusion of the \RSD\ data.} Including DES Y3 data, which provides three times the sky coverage as well as increased depth of field, may provide tighter constraints, though we note that there are some internal inconsistencies in the lens sample galaxies from MagLim and redMaGiC for galaxy clustering and galaxy-galaxy lensing measurements \citep{Abbot/etal:2021,DES:2021bat,Pandey/etal:2022,Elvin-Poole/etal:2023}. Additionally, including non-linear and quasi-non-linear scales might also provide improvement in constraining power, but this would require the ability to model the nonlinear collapse of dark matter. This would require either full N-body simulations or hybrid perturbation-N-body simulations such as COLA simulations \citep[see, e.g.,][]{Tassev/etal:2013} of the TPM model. 

\section{Conclusions} \label{Conclusions}

In this followup analysis to the original TPM analysis performed in B22, we explored the constraining power of using alternative primary CMB anisotropy data in the form of SPT 3G and alternative LSS measurements in the form of RSD measurements from BOSS DR12 and weak lensing measurements from DES Y1. 

We find that SPT 3G primary anisotropy data are in general less constraining of the TPM model than \planck\ anisotropy data, which would allow for the TPM model to resolve the Hubble tension between specifically SPT data and local measurements in the absense of other cosmological information. In particular, we highlight the large shift in the mean value and larger uncertainty of $n_s = 1.003 \pm 0.016$ for the case when TPM is fit to SPT CMB data along with BAO, Supernova, \planck\ CMB lensing, and an $H_0$ prior. This is accompanied by a shift in the amplitude of the transition in the Planck mass (or graviational strength) to $\Omega_0 = -0.072 \pm 0.025$. This shift in the Planck mass is nonzero at 2.9$\sigma$, but we caution that this preference for a shift in $\Omega_0$ is driven by the prior on $H_0$ from local measurements.  

While the TPM model fit to these data results in an increase in $H_0$ to be in better agreement with the $H_0$ prior compared to the \lcdm\ fit to the same data, we note that the preferred values of $S_8$ are slightly higher. When the $H_0$ prior is excluded, the large transitions in the Planck mass and higher $H_0$ preferred values are still allowed by these data at the 95$\%$ confidence level, but these parameter values are found in the non-Gaussian tails of the posterior parameter distributions. The best-fit values are $H_0 = 68.28$ \kmsmpc\ and $\Omega_0 = -0.024$. 

When we additionally include \TTCut, we find that the constraint on the scalar index shifts to $n_s = 0.9746 \pm 0.0057$ and the shift in the Planck mass is constrained to be smaller than 5.1$\%$ at the 95$\%$ confidence level. In this case, the Hubble tension is still alleviated with $H_0 = 70.65 \pm 0.66$ \kmsmpc, though this is helped by a more negative $c_0$ parameter relative to the case with only SPT anisotropy data. From the perspective of the TPM model, shifts in the $c_0$ are less well-motivated because the primary effect of the TPM phenomenology is to change the Planck mass in the early universe. Altering the $c_0$ parameter is interesting, but can be explored more directly using scalar-field models of dark energy in the late universe. Notably, the constraints from this combination are slightly tighter than but still consistent with the constraints from \TTTEEE\ + \base\ + \Hprior.

Again the improvement found by TPM over \lcdm\ is driven entirely by the $H_0$ prior. Without this prior, we find a best-fit value of $H_0 = 68.82$ \kmsmpc, and the amplitude of the transition of the Planck mass is constrained to be smaller than 3$\%$ at the 95$\%$ confidence level. The best-fit value of the amplitude of transition is 0.04$\%$, which is consistent with the \lcdm\ limit of no change in the Planck mass.  

We conclude that large amplitude transitions in the Planck mass result in a suppression of power on small angular scales that are disfavored by SPT and \planck\ data, but SPT data allow for a large shift upward in $n_s$ that can offset this suppression on small scales. The inclusion of the \planck\ anisotropy data restricts the allowed parameter space for $n_s$, and thus limits the TPM model's ability to resolve the Hubble tension. 

One of the main goals of this work was to explore whether the same behavior found by EDE model fits to SPT/ACT data would also be true for the TPM model. In \cite{Smith/etal:2022}, EDE was found to produce higher $H_0$ values without including a prior on $H_0$ when fit to \SPT\ + \TTCut. We find that the TPM model does not also exhibit this behavior. We could constrain the TPM model using ACT primary anisotropy data similar to the EDE fits to ACT both with and without \planck\ anisotropy data. Additionally, while completing this work, we became aware of the public likelihood of SPT data that includes the TT power spectrum \citep{Balkenhol/etal:2022}. We did not use these data sets because the tests using SPT TE and EE data already did not prefer higher values of $H_0$. The different behavior between TPM and EDE is noteworthy because it highlights the fact that not all models that phenomenologically increase the expansion rate prior to recombination to lower the sound horizon and resolve the Hubble tension will fit the available data in a similar way. 

In addition to exploring the ability of SPT data to constrain the TPM model, we also tested the impact of including alternative LSS data. We found that the inclusion of BOSS growth of structure measurements on $f\sigma_8$ resulting from redshift space distortions (RSD) do not significantly alter the allowed TPM parameter space. However, the RSD measurements do restrict the non-Gaussian tail present in the posterior distribution for $S_8$. In B22, we showed that this non-Gaussian tail results from transitions in the Planck mass that occur closer to recombination. 

In addition to growth of structure measurements, we tested the impact of including cosmic shear, galaxy-galaxy lensing, and galaxy clustering measurements from \DES. Because the TPM model does not have a proper description of the non-linear collapse of dark matter, we restricted the \DES\  weak lensing data only include linear scales. We find that these data do not add any significant constraining power when \planck\ primary anisotropy and lensing data as well as BAO and Supernova data are already included. However, the inclusion of \DES\ does slightly restrict the parameter space for $S_8$ similar to the inclusion of \RSD\ data. We conclude that low redshift probes of the matter power spectrum and the growth of structure could constrain the TPM model and particularly the phenomenology of the transition itself. 

Cosmic shear data along with galaxy-galaxy lensing and galaxy clustering from DES Y3, KiDS, and HSC Y3 may also be able to constrain some of the phenomenology of the transition of the TPM model including the rapidity transition and the timing of the transition. However, the data included in this work show little preference for a shift in the Planck mass. 

A potential future effort could consider modifications of the TPM model. However, the data used in this paper are not particularly constraining of $x_T$ or $\sigma$, which  suggests that changing the functional form of the simple step-like transition in the Planck mass is not likely to have a significant effect. One might also explore changing the TPM description of perturbation level parameters as well as details of the scalar-field dynamics, though this has the downside of adding additional parameters. 

We tested the impact of a scalar-field that is minimally coupled to gravity but restricted the equation of state parameter to follow that of the TPM model. We find that this does not greatly affect cosmological parameters. This is because the TPM model scalar-field alone does not have a sufficient energy density prior to recombination to significantly alter background expansion.  

It may be possible to construct a model that has similar phenomenology to TPM or EDE that can mimic the effect of varying the Helium fraction or ionization fraction of electrons so that the reduction in the physical size of the sound horizon at the surface of last scattering does not lead to the same suppression of small scale power through the shift in CMB damping scale. To this point, we strongly caution that a hypothetical new model that can achieve this would likely have multiple new effects relative to the standard \lcdm\ scenario.

\begin{acknowledgments}

This work was performed, in part, under funding from the Jet Propulsion Laboratory, California Institute of Technology, RSA subcontract numbers 1670981 and 1687192 to Johns Hopkins University. This work was carried out using the Advanced Research Computing at Hopkins (ARCH) core facility (rockfish.jhu.edu), which is supported by the National Science Foundation (NSF) grant number OAC 1920103. We would also like to thank Stony Brook Research Computing and Cyberinfrastructure, and the Institute for Advanced Computational Science at Stony Brook University for access to the high-performance SeaWulf computing system, which was made possible by National Science Foundation grant (1531492).  We acknowledge the use of the Legacy Archive for Microwave Background Data Analysis (LAMBDA), part of the High Energy Astrophysics Science Archive Center (HEASARC). HEASARC/LAMBDA is a service of the Astrophysics Science Division at the NASA Goddard Space Flight Center. Finally, we thank Vivian Miranda and Janet Weiland for many helpful suggestions and discussions.

\end{acknowledgments}

\appendix 

\section{Effect of Including an $H_0$ Prior} \label{SPT_H0}

\begin{table*}[!tbp]
\setlength{\tabcolsep}{5pt}
\centering
\hspace{-1cm}
\begin{tabular}{@{}cccccc@{}}

\toprule
& \SPT\                      & \SPT\                     & \SPT\ + \TTCut\  &  \SPT\ + \TTCut\      \\
 & + \base\ + \Hprior\ & + \base\  &   + \base\ + \Hprior\ &  + \base\ \\
\toprule                                                                                                               
$100\theta_{\rm MC}$   & 1.04039 ($1.04042^{+0.00071}_{-0.00072}$)   & 1.03966 ($1.03977^{+0.00075}_{-0.00074}$)  & 1.4091 ($1.04098^{+0.00038}_{-0.00039}$ ) & 1.4073 (1.04072 $\pm$ 0.00036 )  \\
$\Omega_bh^2$          & 0.02271 (0.02277 $\pm$ 0.00030)   & 0.022419 (0.022573 $\pm$ 0.00031)  & 0.02261 (0.02263 $\pm$ 0.00014) & 0.02258 (0.02254 $\pm$ 0.00014)  \\
$\Omega_ch^2$          & 0.1182 (0.1183 $\pm$ 0.0016)   & 0.1176 (0.1182 $\pm$ 0.0015)  & 0.1182 (0.1179 $\pm$ 0.0010) & 0.11834 (0.11856 $\pm$ 0.00099)     \\
$\tau$                 & 0.061 (0.0543 $\pm$ 0.0071)   & 0.0526 ($0.0541^{+0.0069}_{-0.0070}$)  & 0.0485 (0.0516 $\pm$ 0.0075) & 0.0554 ($0.0517^{+0.0073}_{-0.0072}$)        \\
$\ln(10^{10}A_s)$      & 3.049 (3.039 $\pm$ 0.016)   & 3.035 (3.037 $\pm$ 0.015)  & 3.027 (3.034 $\pm$ 0.015) & 3.040 (3.035 $\pm$ 0.015)        \\
$n_s$                  & 1.008 (1.003 $\pm$ 0.016)   & 0.9845 (0.9897 $\pm$ 0.016)  & 0.9733 (0.9746 $\pm$ 0.0057) & 0.9719 (0.9713 $\pm$ 0.0053)     \\
\midrule

$\Omega_0$                  & -0.073 (-0.072 $\pm$ 0.025)   & -0.030 (> -0.100 at 95$\%$)  & -0.013 (> -0.051 at 95$\%$) & -0.0004 (> -0.0301 at 95$\%$)      \\
$x_T$         & -5.06 (-$5.81^{+0.91}_{-0.86}$)   & -5.25 (-$5.57^{+1.04}_{-1.03}$)  & -5.57 ($-5.56^{+1.02}_{-1.01}$) & -6.11 (-$5.32^{+1.14}_{-1.15}$)         \\
$\sigma$         & 1.35 ($1.23^{+0.90}_{-0.84}$)   & 0.76 ($1.34^{+0.98}_{-0.90}$)  & 1.42 ($1.44^{+1.00}_{-0.95}$) & 0.91 ($1.49^{+0.99}_{-0.97}$)        \\

$c_0$              & -0.0111 (> -0.0287 at 95$\%$)   & -0.0051 (> -0.0253 at 95$\%$) & -0.0205 (> -0.0311 at 95$\%$) & -0.0080 (> -0.0228 at 95$\%$)           \\
\midrule
$H_0$                  & 71.87 ($71.94^{+0.86}_{-0.85}$)   & 68.28 ($69.70^{+1.28}_{-1.24}$)   & 70.60 (70.65 $\pm$ 0.66) & 68.82 ($69.09^{+0.69}_{-0.68}$)       \\
$\sigma_8$                  & 0.830 ($0.840^{+0.020}_{-0.021}$)   & 0.810 ($0.832^{+0.020}_{-0.021}$) & 0.825 ($0.839^{+0.018}_{-0.020}$) & 0.814 ($0.831^{+0.019}_{-0.020}$)         \\
$S_8$                  & 0.794 ($0.802^{+0.022}_{-0.023}$)  & 0.812 ($0.820^{+0.023}_{-0.025}$)  & 0.802 ($0.815^{+0.018}_{-0.020}$) & 0.813 ($0.826^{+0.019}_{-0.021}$)        \\
$\chi^2_{\rm Planck TT650TEEE}$ & -   & - & 445.53 & 444.13                 \\
$\chi^2_{\rm Planck low TT}$ & -   & -  & 21.31 & 21.52                \\
$\chi^2_{\rm Planck lowE}$ & -   & -  & 395.73 & 396.03                  \\
$\chi^2_{\rm SPT}$ & 1120.67   & 1118.75  & 1123.46 & 1122.76                    \\
$\chi^2_{\rm CMB \ lensing}$ & 9.17   & 8.98  & 8.37 & 8.69                \\
$\chi^2_{\rm BAO}$ & 6.32   & 5.59  & 8.43 & 5.85                     \\
$\chi^2_{\rm Pantheon}$ & 1035.23   & 1034.73  & 1037.11 & 1034.86                     \\
$\chi^2_{\rm H_0}$ & 0.69   & -  & 5.09 & -                   \\
$\chi^2_{\rm prior}$ & 0.41   & 0.44   & 0.26 & 0.27                   \\
$\chi^2_{\rm tot}$ & 2172.48   & 2168.49  & 3045.28 & 3034.12                    \\
\end{tabular}
\caption{ \label{tab:H0_impact} Maximum Likelihood Parameter Values and, in Parenthesis, Mean plus 68$\%$ Confidence Level Bounds. Each Column Delineates the Likelihood Combination to which the TPM Model is Fit (See Section~\ref{Data} for Details of the Likelihoods). 
}
\end{table*}

\begin{figure*}[!tbp]
\centering

\hspace{-1cm}
\includegraphics[clip,width=\linewidth]{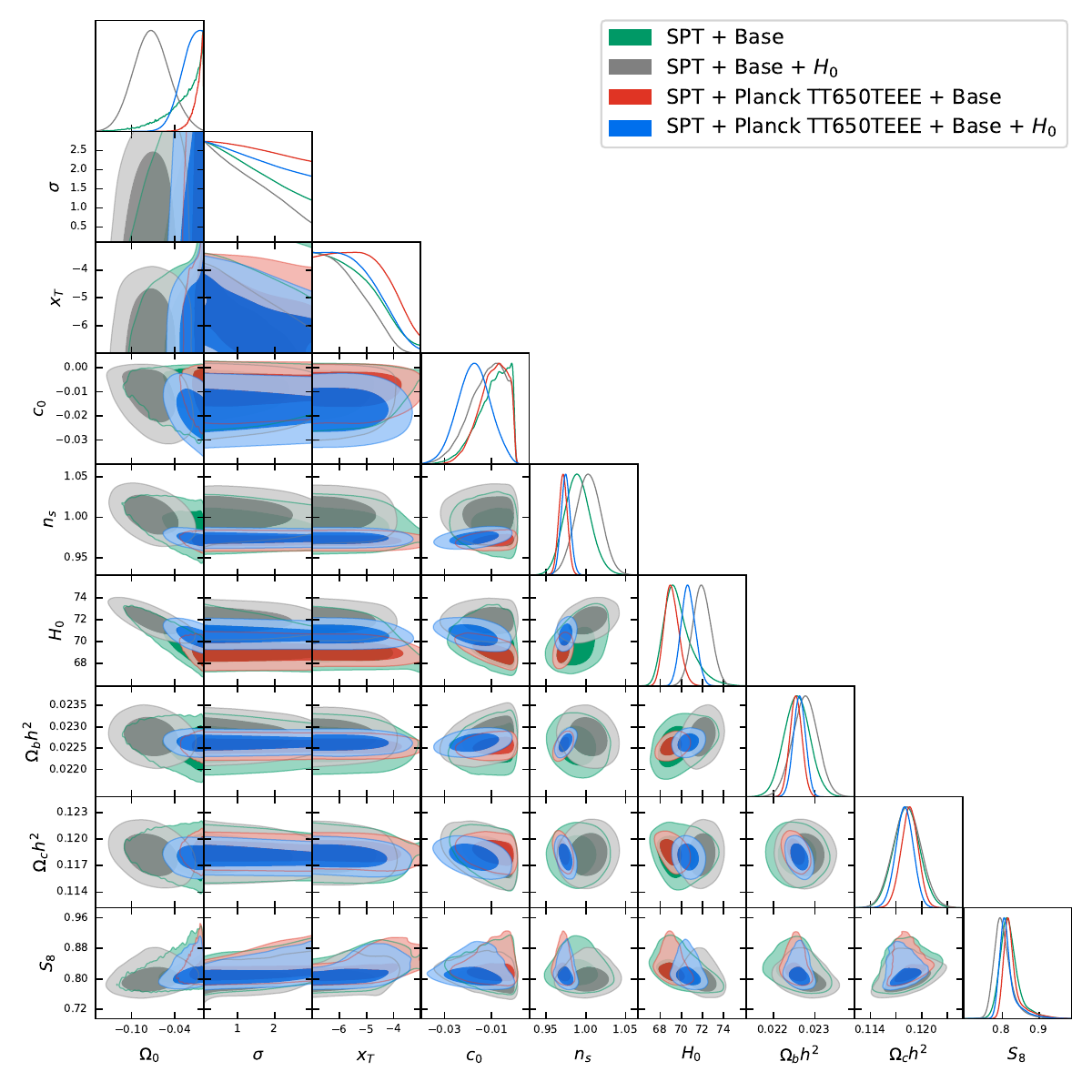}

\vspace{-3mm}

\caption{  Posterior parameter distributions showing the effect of including or excluding the $H_0$ prior from local measurements. The TPM fit to \SPT\ + \base\ (green) does not prefer the same shift in the Planck mass as the case where the $H_0$ prior is included (gray); however, the posterior distribution has a non-Gaussian tail that allows for larger amplitude transitions. When \TTCut\ is included, the allowed parameter space for $\Omega_0$ is restricted for both the case where the $H_0$ prior is included (blue) or excluded (red). Without the $H_0$ prior, the $c_0$ shifts upward (less negative). This corroborates the assertion that the shift in $c_0$ for the TPM fit to \SPT\ + \TTCut\ + \base\ + \Hprior\ is driven by the prior on $H_0$ and the limitation of the $\Omega_0$ parameter to increase $H_0$ when SPT and \planck\ data are combined. We cut the parameter values for $\Omega_0 > 0.0035$ for visualization purposes.
}

\label{fig:H0_vs_NoH0}
\end{figure*}

In this appendix, we explore the impact of the $H_0$ prior on the fits to \SPT\ + \base\ + \Hprior\ and \SPT\ + \TTCut\ + \base\ + \Hprior. We show the results of these tests in Figure~\ref{fig:H0_vs_NoH0} and Table~\ref{tab:H0_impact}. We find that for both the fits to \SPT\ + \base\ + \Hprior\ and \SPT\ + \TTCut\ + \base\ + \Hprior, the mean value of $H_0 < 70$ \kmsmpc. For the TPM fit to \SPT\ + \base\ + \Hprior, the constraint is $H_0 = 69.70^{+1.28}_{-1.24}$ \kmsmpc; however, the parameter distribution function is non-Gaussian as evidenced by the posteriors shown in Figure~\ref{fig:H0_vs_NoH0} and by the best-fit value of $H_0 = 68.28$ \kmsmpc. This means the mean value is skewed to higher values by the presence of a non-Gaussian tail. 

Additionally, there is a non-Gaussian tail for the amplitude of the transition in the Planck mass, $\Omega_0$, whose best-fit value corresponds to 2.4$\%$ shift in the Planck mass, but is constrained at the 95$\%$ level to have a transition smaller than 10$\%$. The $c_0$ parameter, which acts like a scalar-field equation of state parameter in the late universe, is consistent with a value of 0. This corresponds to the TPM f(R) case. 

Importantly, without the $H_0$ prior, the constraint on $n_s = 0.9897 \pm 0.016$, which is lower than the constraint $n_s = 1.003 \pm 0.016$ when the prior is included. This highlights that the $H_0$ prior is driving this shift in $n_s$ just as it drives the increase in the preferred value of $H_0$ and decrease in preferred value of $\Omega_0$.

When \TTCut\ data are added to this likelihood combination, the data show no preference for a shift in the Planck mass with a best-fit value for the amplitude of the transition being 0.04$\%$, which is consistent with no transition in the Planck mass (i.e. the \lcdm\ limit). The amplitude is constrained to be less than $3\%$ at the $95\%$ confidence level. This highlights that these data show no preference for the core phenomenology of the TPM model.

\section{Comparison of the TPM Versus TPM f(R) Models for SPT Data} \label{SPT_fR}

\begin{figure*}[!tbp]
\centering

\hspace{-1cm}
\includegraphics[clip,width=\linewidth]{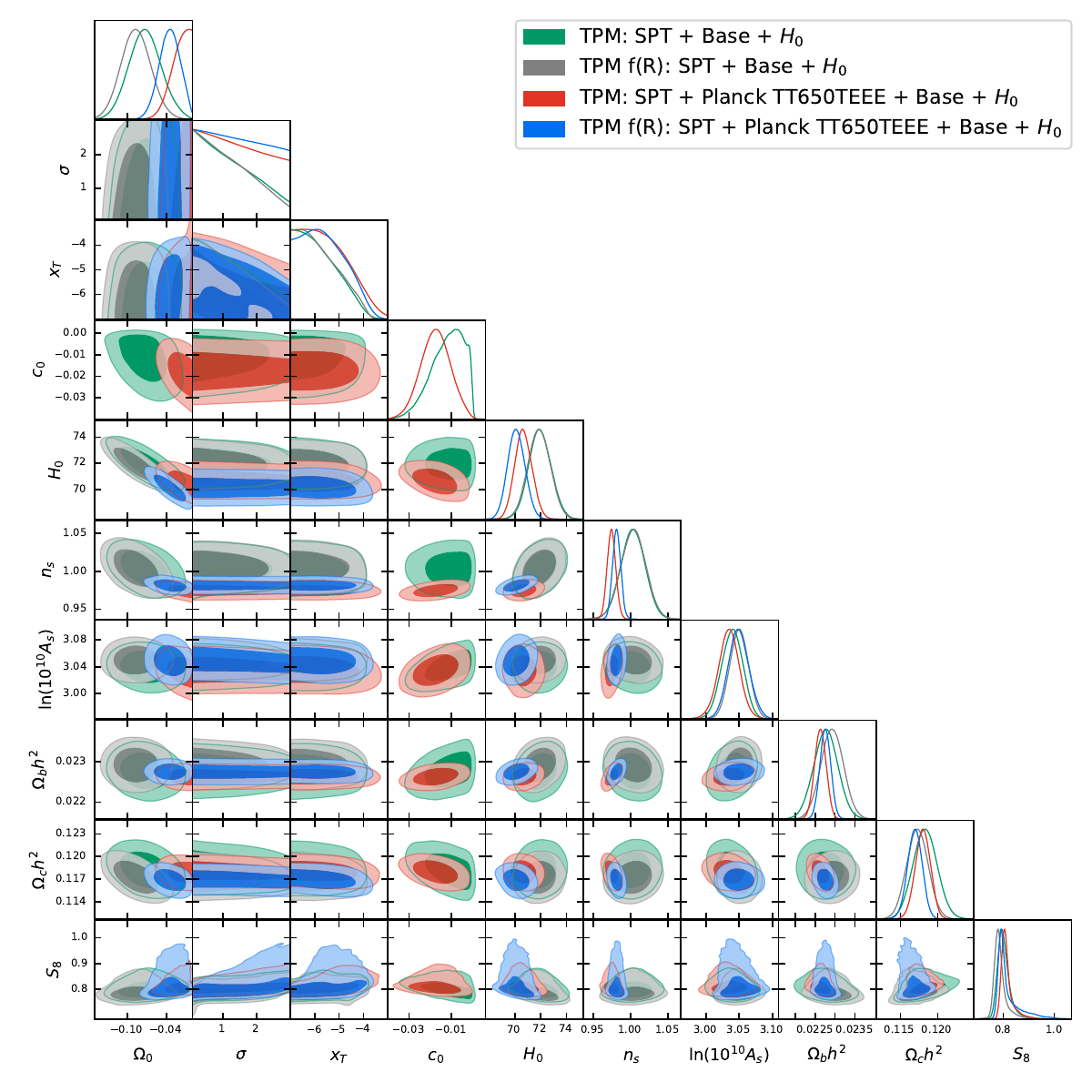}

\vspace{-3mm}

\caption{Posterior distribution functions for the TPM and TPM f(R) model fits to \SPT\ + \base\ + \Hprior\ and \SPT\ + \TTCut\ + \base\ + \Hprior. The TPM f(R) model is the subset of the TPM model where the $c_0$ is fixed to 0, which isolates the phenomenology of the transition of the Planck mass. The $c_0$ parameter acts as a equation of state parameter for the scalar-field in the late universe. For the fits to \SPT\ + \base\ + \Hprior, there is little difference between the TPM (green) and TPM f(R) (gray) models because the data allow for large shifts in the Planck mass. When the \TTCut\ data are added, this model freedom is removed and the full TPM model (red) shifts $c_0$ to be more negative to increase $H_0$. The TPM f(R) model (blue) can still achieve higher $H_0$ values, but the fit to data is worse. We cut the parameter values for $\Omega_0 > 0.0035$ for visualization purposes.
}

\label{fig:TPM_vs_TPM_fR}
\end{figure*}

\begin{table*}[!tbp]
\setlength{\tabcolsep}{5pt}
\centering
\hspace{0cm}
\begin{tabular}{@{}cccccc@{}}

\toprule
& TPM                       & TPM f(R)                        & TPM   &  TPM f(R)    \\
 & \SPT\  &  \SPT\ & \SPT\ + \TTCut\ & \SPT\ + \TTCut\    \\
 & + \base\ + \Hprior\ &  + \base\ + \Hprior\ & + \base\ + \Hprior\ &  + \base\ + \Hprior\    \\
\toprule                                                                                                               
$100\theta_{\rm MC}$   & 1.04039 ($1.04042^{+0.00071}_{0.00072}$)   & 1.04086 (1.04067 $\pm$ 0.00069)  & 1.4091 ($1.04098^{+0.00038}_{-0.00039}$) & 1.04124 (1.04133 $\pm$ 0.00039 )  \\
$\Omega_bh^2$          & 0.02271 (0.02277 $\pm$ 0.00030)   & 0.022939 (0.022928 $\pm$ 0.00028)  & 0.02261 (0.02263 $\pm$ 0.00014) & 0.02273 ($0.02276^{+0.00014}_{-0.00013}$)  \\
$\Omega_ch^2$          & 0.1182 (0.1183 $\pm$ 0.0016)   & 0.1174 (0.1173 $\pm$ 0.0014)  & 0.1182 (0.1179 $\pm$ 0.0010) & 0.11732 ($0.11691^{+0.00097}_{-0.00096}$)     \\
$\tau$                 & 0.061 (0.0543 $\pm$ 0.0071)   & 0.0565 (0.0558 $\pm$ 0.0070 )  & 0.0485 (0.0516 $\pm$ 0.0075) & 0.0550 (0.0588 $\pm$ 0.0074)        \\
$\ln(10^{10}A_s)$      & 3.049 (3.039 $\pm$ 0.016)   & 3.052 (3.050 $\pm$ 0.014)  & 3.027 (3.034 $\pm$ 0.015) & 3.041 (3.049 $\pm$ 0.015)        \\
$n_s$                  & 1.008 (1.003 $\pm$ 0.016)   & 1.004 (1.004 $\pm$ 0.016)  & 0.9733 (0.9746 $\pm$ 0.0057) & 0.9800 (0.9814 $\pm$ 0.0054)     \\
\midrule

$\Omega_0$                  & -0.073 (-0.072 $\pm$ 0.025)   & -0.088 (-0.086 $\pm$ 0.023)  & -0.013 (> -0.051 at 95$\%$) & -0.034 (> -0.067 at 95$\%$)      \\
$x_T$         & -5.06 (-$5.81^{+0.91}_{-0.86}$)   & -6.14 (-$5.79^{+0.93}_{-0.87}$)  & -5.57 (-$5.56^{+1.02}_{-1.01}$) & -4.98 (-$5.60^{+0.97}_{-0.95}$)         \\
$\sigma$         & 1.35 ($1.23^{+0.90}_{-0.84}$)   & 1.16 ($1.19^{+0.86}_{-0.81}$)  & 1.42 ($1.44^{+1.00}_{-0.95}$) & 0.25 ($1.49^{+0.99}_{-0.97}$)        \\

$c_0$              & -0.0111 (> -0.0287 at 95$\%$)   & 0 (fixed) & -0.0205 (> -0.0311 at 95$\%$) & 0 (fixed)          \\
\midrule
$H_0$                  & 71.87 ($71.94^{+0.86}_{-0.85}$)   & 71.98 ($71.88^{+0.86}_{-0.85}$)   & 70.60 (70.65 $\pm$ 0.66) & 69.90 (70.12 $\pm$ 0.67)       \\
$\sigma_8$                  & 0.830 ($0.840^{+0.020}_{-0.021}$)   & 0.816 ($0.827^{+0.018}_{-0.019}$) & 0.825 ($0.839^{+0.018}_{-0.020}$) & 0.804 ($0.840^{+0.044}_{-0.035}$)         \\
$S_8$                  & 0.794 (0.802 $\pm$ 0.027)  & 0.777 ($0.788^{+0.019}_
{-0.021}$)  & 0.802 ($0.815^{+0.018}_{-0.020}$) & 0.787 ($0.820^{+0.043}_{-0.036}$)        \\
$\chi^2_{\rm Planck TT650TEEE}$ & -   & - & 445.53 & 450.78                 \\
$\chi^2_{\rm Planck low TT}$ & -   & -  & 21.31 & 20.65                \\
$\chi^2_{\rm Planck lowE}$ & -   & -  & 395.73 & 395.95                  \\
$\chi^2_{\rm SPT}$ & 1120.67   & 1122.48  & 1123.46 & 1123.55                    \\
$\chi^2_{\rm CMB \ lensing}$ & 9.17   & 9.73  & 8.37 & 11.39               \\
$\chi^2_{\rm BAO}$ & 6.32   & 5.53  & 8.43 & 5.70                     \\
$\chi^2_{\rm Pantheon}$ & 1035.23   & 1034.73  & 1037.11 & 1034.74                    \\
$\chi^2_{\rm H_0}$ & 0.69   & 0.51  & 5.09 & 9.25                   \\
$\chi^2_{\rm prior}$ & 0.41   & 0.45   & 0.26 & 0.05                   \\
$\chi^2_{\rm tot}$ & 2172.48   & 2173.43  & 3045.28 & 3052.04                    \\
\end{tabular}
\caption{ Maximum Likelihood Parameter Values and, in Parenthesis, Mean plus 68$\%$ Confidence Level Bounds. Each Column Delineates the model and Likelihood Combination (See Section~\ref{Data} for Details of the Likelihoods). \label{tab:fR}
}
\end{table*}

We have explored constraints on the TPM model with the $c_0$ parameter free to vary. However, it is also interesting to explore what happens when this parameter is fixed to 0 because this isolates the core phenomenology of the TPM model (i.e. the shift in the Planck mass). The $c_0$ parameter effectively acts as the equation of state parameter for the scalar-field in the late universe. More negative values of $c_0$ lead to larger values of $H_0$ similar to how more negative values of $w$ in $wCDM$ models tend to lead to larger preferred values of $H_0$. We refer to this case where $c_0 = 0$ as the TPM f(R) model because this model exists within the broad class of f(R) modified gravity models. 

In this appendix, we explore the constraints on the TPM f(R) model. We run MCMCs for the TPM f(R) model fits to \SPT\ + \base\ + \Hprior\ and \SPT\ + \TTCut\ + \base\ + \Hprior, and show the results in Figure~\ref{fig:TPM_vs_TPM_fR} and Table~\ref{tab:fR}. 

There is little difference between fitting the TPM and TPM f(R) models to \SPT\ + \base\ + \Hprior\ because these data can compensate large shifts in the Planck mass by increasing $n_s$. The full TPM model has the additional model freedom of the $c_0$ parameter to increase $H_0$. Thus, the \SPT\ + \base\ + \Hprior\ prefer more negative values of $\Omega_0$ in the TPM f(R) model than the full TPM model. 

The TPM f(R) model fits the \SPT\ + \base\ + \Hprior\ slightly worse than the TPM model, but the additional model freedom from varying $c_0$ does not significantly improve the fit to data. 

When \TTCut\ data are additionally included, the $n_s$ parameter is more tightly constrained, which in turn restricts by how much the $\Omega_0$ parameter can shift and still fit the data well. Nevertheless, the $H_0$ prior forces the model to find parameters that have higher $H_0$
values. Therefore, in the full TPM model, the $c_0$ parameter shifts to more negative values and the best-fit value for the shift in the Planck mass is only $1.3\%$. 

For the TPM f(R) model, the best-fit value for the shift in the Planck mass is 3.4$\%$, and the best-fit value for $H_0 = 69.90$ \kmsmpc. Importantly, the TPM f(R) model fits the 
\SPT\ + \TTCut\ + \base\ + \Hprior\ significantly worse than the full TPM model with a $\Delta \chi^2 = 6.76$ for only one additional parameter.

\bibliographystyle{aasjournal}
\bibliography{cosmology}

\end{document}